\documentclass[12pt]{article}
\usepackage{enumitem}
\usepackage[english]{babel}
\usepackage[utf8]{inputenc}
\usepackage[comma,authoryear]{natbib}
\usepackage{amsmath,multibib}
\usepackage{amssymb,amsthm}
\usepackage{mathrsfs}
\usepackage{graphicx}
\usepackage{url}
\usepackage{caption,subcaption}
\usepackage{booktabs}
\usepackage{multirow}
\usepackage{epsfig,float}
\usepackage[usenames,dvipsnames]{color} 

\usepackage{mathrsfs,bm} 
\newcommand{\CUST}{\operatorname{CUST}}

\newcommand{\rel}{\operatorname{rel}}

\theoremstyle{plain}
\newtheorem{thm}{Theorem}

\newcommand{\bthm}{\begin{thm}}
\newcommand{\ethm}{\end{thm}}
\newcommand{\bpf}{\begin{proof}}
\newcommand{\epf}{\end{proof}}
\theoremstyle{definition}

\newtheorem{rem}{Remark}
\usepackage[margin=.9in,a4paper]{geometry}
\usepackage[compact]{titlesec}
\numberwithin{equation}{section}
\usepackage{deep-macro}
 
\usepackage{epigraph}
\usepackage{relsize}

\usepackage{listings}
\usepackage{tabularx}
\newcolumntype{Y}{>{\centering\arraybackslash}X}
\newcolumntype{f}{>{\centering\arraybackslash}X}
\newcolumntype{h}{>{\hsize=.5\hsize\centering\arraybackslash\extracolsep{.1em}}X}

\newcolumntype{C}[1]{>{\hsize=#1\hsize\centering\arraybackslash}X}

\begin{document}
\vskip3em
\begin{center} 
{\Large {\bf On The Problem of Relevance in Statistical Inference} }
\end{center}
\vskip.1in
\begin{center}
\begin{tabularx}{.97\linewidth}{ff}
{\bf Subhadeep Mukhopadhyay}\footnote{Correspondence should be sent to  deep@unitedstatalgo.com.}& {\bf Kaijun Wang}\\
\texttt{deep@unitedstatalgo.com} & \texttt{kwang2@fredhutch.org}
\end{tabularx}
\end{center}

\begin{abstract}
Given a large cohort of ``similar'' cases one can construct an efficient statistical inference procedure by learning from the experience of others (also known as ``borrowing strength'' from the ensemble). But what if, instead, we were given a massive database of \emph{heterogeneous} cases? It's not obvious how to go about gathering strength when each piece of information is fuzzy. The danger is that, if we include irrelevant cases, borrowing information might heavily damage the quality of the inference! This raises some fundamental questions for big data inference: When (not) to borrow? Whom (not) to borrow? How (not) to borrow? These questions are at the heart of the ``Problem of Relevance'' in statistical inference -- a puzzle that has remained too little addressed since its inception nearly half a century ago \citep{efron1972EB, mallows1982, efron2019}. This paper develops a \emph{new model} of large-scale inference to tackle some of the unsettled issues that surround the relevance problem. Through examples, we will demonstrate how our new statistical perspective answers previously unanswerable questions in a realistic and feasible way.


\end{abstract}
\linespread{1.28}
\renewcommand{\baselinestretch}{1.34}
\setlength{\parskip}{1.4ex}
\section{Introduction}
\begin{figure}[ht]
        \centering
        \includegraphics[width=.68\linewidth,trim=1cm 0cm 1cm .5cm]{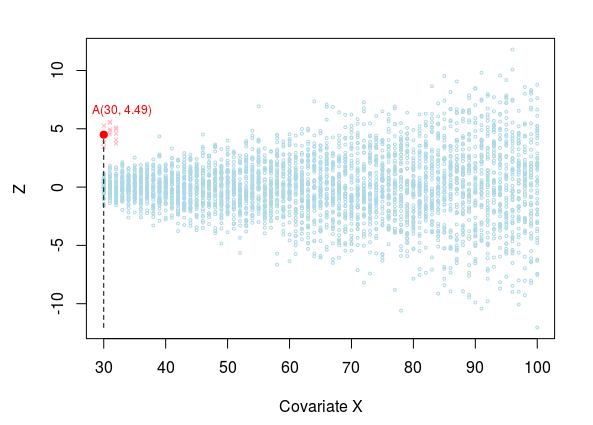}
        \vspace{-.5em}
        \caption{The \texttt{funnel} problem: $z_i\sim \cN(\te_i,\si_i^2)$, $i=1,\ldots,N=3,565$ where the variability is increasing linearly as a function of $x$: $\si(x_i)= x_i/21-0.71$, $30 \le x_i \le 100$. For each $x$ between $30$ and $100$, we have $50$ $z$-values with $\te_i=0$. Additional $15$ true signals ($5$ at each locations $x=30$ to $x=32$) with $\te_i=4.49$ are indicated by the light red color; they are buried in noisy background fluctuations. Obviously, the data generating process (relationship between $z$ and $x$) will be considered as unknown in our analysis. The red dot is the target case A with $z_A=4.49$ and $x_A=30$, for which we like to perform customized inference in a completely \textit{nonparametric} manner. Supplementary A discusses some of the practical motivation behind this \texttt{funnel} data problem.
        } \label{fig:funnel}
    \end{figure}

We are interested in the following question: Given a large number of summary statistics $z_1,\ldots z_N$ from $N$ cases (genes, voxels, neurons, patients, customers, baseball players, etc.) how to efficiently perform customized inference (testing as well as estimation) for a particular individual case? If we assume that each $z_i$ is \textit{equally} informative or relevant to the case in hand, a precise individualized-inference can be delivered by learning from the experience of others \citep{efron1972EB, mallows1982, efron2010book, efron2019}. However, this assumption of ``uniformity of relevance'' breaks down when dealing with large assembly of \textit{heterogeneous} cases,  something that is becoming a norm in almost all modern data-science applications including neuroscience, genomics, healthcare, and astronomy. 
\vskip.5em
{\bf Origin of the Relevance Problem}. To illustrate this point, consider the following example: where for each of the $N=3,565$ cases we are given a z-score $z_i$ and an extra piece of information in the form of a covariate $x_i$ (e.g., location information of voxels, genomic biomarker of patients, playing position pitcher/nonpitcher of baseball players, etc.) that captures the domain-context. We seek to perform an inference for the target case A (the red dot in Fig.\ref{fig:funnel}) by taking its characteristic feature $x_A=30$ into account. 
    
\begin{figure}
    \centering
      \includegraphics[width=.7\linewidth,trim=1cm 0cm 1cm 1cm]{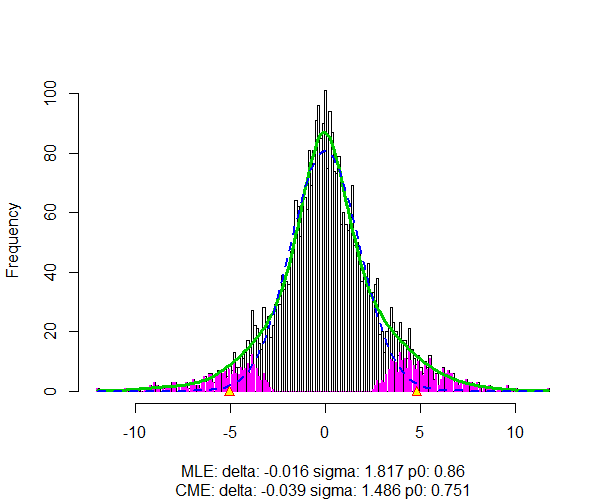}\\[2em]
    \includegraphics[height=.53\linewidth,trim=1cm 0cm 1cm .5cm]{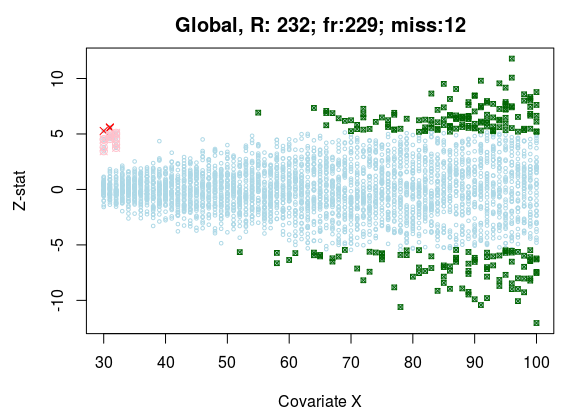}
   \caption{Heterogeneity blankets the true signals (in light red color), and make them invisible from the global reference frame. The background variability makes the noises look ``bigger'' than the signals! As a result, all global large-scale inference methods mostly end up selecting loud noises. Here we display the result of local false discovery method \citep{efron08} that misses $12$ out of $15$ true signals and picks $229$ false ones (green colored)--acts as a noise amplifier instead of a signal detector. Similar things happen for other methods---see Supplementary Fig. \ref{supp:comp}.} \label{fig:global_simu}
\end{figure}

{\bf Significant? But, Relative to What?} Let's start with the most basic question: whether case A $(x_A=30,z_A=4.49)$ is statistically significant, or at least intriguing enough to study in detail. However, the word ‘significance’ only makes sense if we know \textit{relative to what}? A declaration of statistical significance is not an absolute verdict; it's a relativistic concept that depends on what we consider as the reference or baseline. The conventional practice adopts the ensemble of aggregated observations as the `fixed' relevant comparison set  (an ``absolute'' frame of reference) for each individual case. This global one-size-fits-all strategy leads to troublesome results, as is visible in Figure \ref{fig:global_simu}.
 
   \begin{quote}
\vspace{-.53em}
{\small \textit{``The
relevance rule of `all the cases that show up together on my desk' doesn't stand up to scrutiny, but formulating an alternative seems difficult}.}'' (Efron, 2019) \nocite{efron2019}
\vspace{-.25em}
\end{quote} 
  
In hindsight, it is no wonder that heterogeneity makes it silly to compare a specific case with the whole population---the comparison has to be done in relation to `something else' other than the ensemble. But, what is that something else? What other alternatives we have? Should we instead compare with the cases that share exactly same characteristics (i.e, use $55$ z’s with $x= 30$ for case A)?
Clearly, this is not a wise decision, since it produces too little direct data (`$N > 1000$ is necessary,' \citealp{efron2008com}) to deliver any reliable large-scale inference result.

{\bf Empirical Bayes and Relevant Prior}. The relevance problem also arises when we want to estimate the actual effect-size of a non-null case. Instead of a point estimate, researchers often prefer 
the whole posterior uncertainty distribution (i.e., the probability distribution of all possible values given the actual data) of the associated parameter. The main hurdle to realizing this goal is the `prior,' which needs to be estimated in an objective manner before we apply Bayes' theorem.  Traditional empirical Bayes \textit{learns} the global prior (`fixed' for all cases) from the full sample $z_1,\ldots, z_N$ \citep{efron2016, deep18nature}. But the practical concern here is whether the global prior is relevant for the case in hand. Surely it would be if we had one grand m{\'e}lange of homogeneous observations, which, unfortunately, is not the case in most practical problems. Therefore, it is natural to ask `which others' carry relevant information for case A. To accurately learn such a customized prior, we need hundreds or even thousands of parallel samples that are related to case A--an impractical expectation. That said, the question remains: how to design a justified recipe for estimating an individualized relevant prior?
The answer to this question holds an important key to the practice of empirical Bayes in the era of big heterogeneous datasets.

\begin{figure}[t]
        \centering
\includegraphics[width=.9\linewidth,trim=2cm 1cm 2cm 1cm]{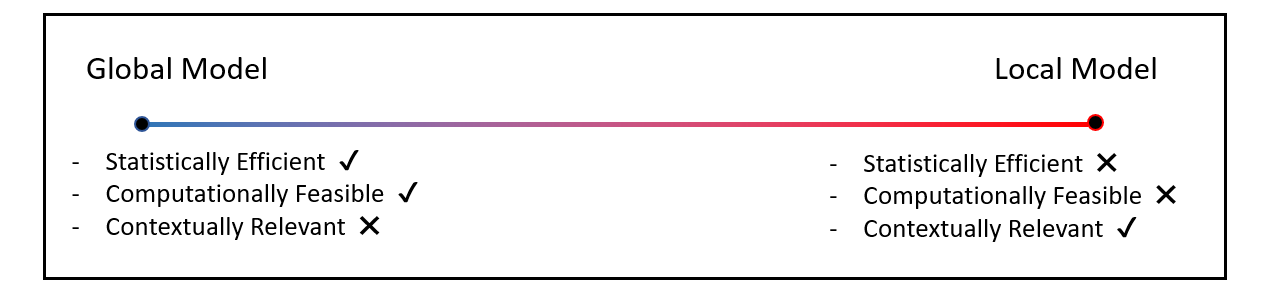}
\vskip.5em
\caption{The relevance paradox: a classic ``Catch-22'' situation.}\label{fig:relpara}
\vspace{-.25em}
    \end{figure}

\vskip.5em
{\bf The Relevance Paradox}. It is evident from the discussions so far that big data inference (both simultaneous testing and estimation) poses some unique practical challenges: on the one hand the full-data-based global models are statistically efficient but not contextually relevant; on the other hand, the local inferential models are either uncalculable or absurdly noisy.  Figure \ref{fig:relpara} depicts this bizarre quagmire, which shows that both global and local modes of inferences are  unfruitful avenues for harnessing heterogeneous large datasets.  
So, can we find an algorithmic solution to reconcile this seemingly paradoxical situation emerging from the relevance problem? The ``ideal'' scenario would be to have a  customized-inference framework (in between two extremes: global and local) that is contextually relevant and at the same time sacrifices very little, if any, efficiency. Currently, there exist no such  practical theory or implementation protocols that can come close to this much sought-after goal of simultaneously improving the quality and relevance of statistical inferences across the cases.

\vskip.5em

{\bf Goals and Contributions}. So, where do we stand now? Over the past several decades, great progress has been made in the field of `large-scale inference'  that helped to create a vast and impressive inventory of global inference methods; see the monographs \cite{efron2010book}, \cite{efron2016computer}, and references therein. However, as these methods are primarily useful for large homogeneous problems, the question arises as to how to modify them in order to make them applicable for real-world large \textit{fuzzy} datasets? Can we develop a \textit{general mechanism} that can ``convert'' these global inference algorithms into individualized ones? These are some of the questions that motivated us to embark on this research, which has three interconnected parts:

~~(1) \textit{Diagnosis}: Given $\{(x_i,z_i)\}_{i=1}^N$ how can we check \textit{whether} the global analysis is valid or not for $x=x_0$ cases? Can we develop a nonparametric diagnostic tool?

~~(2) \textit{Modeling}: If the global assumption is unreasonable, then the next question is \textit{how} the local $z$-values at $x_0$ are different from the aggregated $z_1,.\ldots,z_N$. For example, as we can see from Fig. \ref{fig:funnel} the distributional characteristics of the local $z$-values at $x=30$ are very different from the ensemble one. In fact, the marginal variance is almost $3$ times that of the variance of the $z$-values at $x=30$. The concept of ``relevance'' function, as we will soon see, critically depends on the difference (``deviance'') between the local and global distributions.

~~(3) \textit{Synthesis}: Finally, the question is: how to ``sharpen'' the 
aggregated messy $z_1,\ldots,z_N$ to produce a relevant comparison set?  Can we do it in a fully automated and data-driven manner that works even for multivariate ${\bf x}$? If our specially-designed dummy $z$-values faithfully capture the \textit{distributional heterogeneity} (intrinsic uncertainty and fluctuations) at $x_0$, then we can use them for ``borrowing strength.'' We accomplish this goal by synthesizing \texttt{LASERs}--Artificial RELevant Samples. They provide a direct ``one-shot'' approach to convert \textit{any} global inferential method into a customized one. A schematic representation of the algorithmic workflow is given below:
\vskip.65em
~\fbox{Global inference engine} + \fbox{$N$-laser samples at $x_0$} \,= \,\fbox{Tailor-made inference for $x_0$ cases}
\vskip.5em
This simple modular architecture hugely simplifies the implementation of our approach. Since, we can now utilize all the existing global inference algorithms (and associated R-routines) to produce its individualized versions. But to get there, we first have to introduce some modern nonparametric concepts and notation, which will lay the basis for a statistical theory of relevance. This is done in Section \ref{sec:theory}. The key ideas discussed here are: relevance function, global-to-local conditional density representation, and a computational recipe for generating LASERs. In Section \ref{sec:ci}, we provide a complete picture of LASER-guided customized inference, specifically touching upon the significance of relevance in microinference, empirical Bayes, and reproducible inference. Sections \ref{sec:app1} and \ref{sec:app2} deal with two real-applications: DTI neuroscience data and kidney data. We end with some final remarks in the last section. The Supplementary Appendix contains additional details. 

\vskip.25em
{\bf Notation}. By $F_Z(z) = \Pr(Z \le z)$ we denote the marginal cumulative distribution function (cdf) of the random variable $Z$, while $Q_Z(u),\,0<u<1$ is the the respective quantile function. We drop the subscripts, whenever it is clear from the context. Note that for $Z$ continuous $Q(u)$ is simply $F^{-1}(u)$ for $0<u<1$. We will denote conditional cdf $\Pr(Z \le z|X=x)$ by $F(z|x)$. The marginal and conditional densities (pdfs) are respectively expressed as $f(z)$ and $f(z|x)$. The empirical cdfs are denoted by $\wtF(z)$ and $\wtF(z|x)$. The sets $\bbZ$ and $\bbZ_x$ contain z-values for the full data and target group with cases $X=x$. Finally, $\Rel(\bbZ,\bbZ_x)$ stands for relevance of the full data for the group with $X=x$.






 

 






\vspace{-.25em}
\section{A Statistical Theory of Relevance} \label{sec:theory}
So far we have handled the issue of relevance in an informal way. However, any serious progress in this direction, will first and foremost, require a mathematically precise statistical description of what we mean by `relevance.' This section is organized with this goal in mind: to introduce the fundamental modeling principles that are needed to establish a general theory of relevance.
\subsection{The Relevance Function}
The question of relevance $\Rel(\bbZ,\bbZ_x)$
is intimately tied with the question of representativeness: How representative is the full data for a target group with feature $X=x$? If the statistical (distributional) characteristics of the target z-values $\bbZ_x$ differ significantly from the statistical characteristics of the ensemble $\bbZ$, then there is a high risk of getting erroneous results using global inferential methods. Therefore, it makes sense to define relevance $\Rel(\bbZ,\bbZ_x)$ as ``information sharing'' between the global marginal $Z$ and local conditional $Z|X=x$, which can be measured by understanding how close (or different) the shape of $f(z|x)$ is to $f(z)$. To formalize the idea, consider the ratio
\[\dfrac{f(z|x)}{f(z)},\]
which captures the ``amount'' of information overlap (or relevance) between the combined data and cases with $X=x$. We rewrite this ratio (which is a general function of $z$) in the quantile-domain by substituting $F(z)=u$ in the previous expression: 
\beq \label{eq:cdu} d_x(u) \,:=\, d(u;Z,Z|X=x)\,=\,\dfrac{f(Q(u)|x)}{f(Q(u))},~~\text{for}~ 0 \le u \le 1. \eeq
to make it a proper density function over the unit interval, since $$\int_0^1 d(u;Z,Z|X=x)  \dd u=1.$$ We now formally define $d_x(u)$ as the relevance function (or kernel) that compares the distribution of the marginal $Z$ with that of $Z|X=x$; this also justifies its notation.
\begin{rem}[Shape of $d_x$ under homogeneity]
If $Z \perp X$, i.e., when $x$ contains no relevant information for $z$, $d_x(u)$ reduces to the uniform density
\[d_x(u) = 1,~~0<u<1~~(\text{for all $x$}).\]
Since under independence: $f(z|x)=f(z)$ for all $x$. This is further elaborated in Section \ref{sec:mrel}.
\end{rem}

\begin{rem}[Special case when $X$ is binary]
Suppose we are dealing with a scenario where we have two groups (baseball batter vs. pitcher or left vs. right-half brain voxels etc.) indicated by $X=0$ and $X=1$. In this simplistic scenario, the relevance function $d_x(u)$ reduces to
\beq 
d_x(u) = \dfrac{f(Q_Z(u)|X=1)}{f(Q_Z(u))} = \dfrac{\Pr(X=1|Z=Q_Z(u))}{\Pr(X=1)},
\eeq
where the last equality follows from Bayes theorem. This suggests relevance function is proportional to the conditional probability of group assignment given $Z=z$. This is also known as the `propensity score function,' which is used for controlling selection biases in observational studies. The alternative interpretation of our relevance function as a propensity-weighing function (in the context of binary $X$) was conjectured by an erudite reviewer, whom we sincerely thank.

However, it is important to keep in mind that for real-world problems, we \textit{don't know} the right groups with comparable cases. The challenge lies in empirically deducing the relevance law $d_x(u)$ in a way that is computationally efficient and works for multivariate features ${\bf X}$.
\end{rem}
\subsection{The Global-to-Local Representation} \label{sec:g2lm}
One can recover conditional density through the relevance function using the following universal recipe:
\beq \label{eq:g2l} f(z|x)\,=\,f(z) \times d\big(F(z); Z,Z|X=x\big). \eeq 
Justification of this representation immediately follows from the fact that $d(F_Z(z); Z,Z|X=x)$ is simply $f(z|x)/f(z)$, by virtue of the definition \eqref{eq:cdu}. For brevity's sake, we will refer $d(F_Z(z);$ $Z,Z|X=x)$ as $d_x(z)$ throughout the article.
\vskip.5em
{\bf Interpretations}. Our two-component conditional density decomposition formula \eqref{eq:g2l} can be interpreted from several angles: 
\vskip.2em
$\bullet$ We call it `global-to-local' since it admits the following decomposition: 
\[ \text{local $f(z|x)$~=~ global $f(z)$ $\times$ ``relevance correction'' as a function of $x$ and $z$.} \]
Hence, local distributions can be created by \textit{warping} the shape of the global distribution via $d_x(z)$. This allows us to ``borrow strength from the ensemble'' for efficient modeling.

\vskip.2em
$\bullet$ By its construction, the relevance function extracts all the `fine details' that are \textit{exclusive} to $Z|X=x$, i.e., different from the marginal $Z$. Accordingly,
the shape of $d_x(z)$ contains important clues about the degree of \textit{required customization} to go from $f(z)$ to $f(z|x)$. We will elaborate more on this in Section \ref{sec:mrel}.

\vskip.2em
$\bullet$ Another  noteworthy aspect of the global-to-local representation lies in expressing the relevance function $d \hspace{-.06em}\circ \hspace{-.06em} F(z)$ in the quantile or rank-transform domain, i.e., expressing it as a function of the probability integral transform $F(Z)$. This allows a \textit{robust} way to construct the local conditional distribution from the global marginal. 

\vskip.5em

{\bf Polynomials of Ranks}. Perform a 
robust and efficient nonparametric estimation of the relevance function $d \hspace{-.06em}\circ \hspace{-.06em} F(z)$ by expressing it as a linear combination of polynomial of rank transformation $F(z)$. For $Z$ continuous (which is the case here), one can easily construct such polynomials of rank transformation $F(z)$, according to the following recipe\footnote[2]{The general case where $Z$ is mixed (either discrete or continuous) is discussed in \cite{DeepLPKsample19}, \cite{D20copula} and \cite{Deep20USA}.}: standardize $F(z)$ by its mean $\Ex[F(Z)]=1/2$ and variance $\Var[F(Z)]=1/12$ to get 
\[T_1(Z;F)= \sqrt{12}\big(F(Z) - 1/2\big).\]
Construct an orthonormal basis $\{T_j(Z;F)\}_{j \ge 1}$ for $\sL^2(F)$ by applying Gram-Schmidt orthonormalization on the set of functions $\{T_1,T_1^2,\ldots\}$.

\vskip.5em
{\bf Empirical Rank-Polynomials}. However, we cannot directly construct these polynomials, as they depend on the unknown $F_Z$. Given $z_1,\ldots,z_N$ we construct empirical polynomials $\{T_j(Z;\wtF)\}_{j \ge 1}$ which by design obey the following property:
\[\int_z  \wtT_j(z) \dd \wtF(z) \,=\,0 ~\,\text{and}~\int_z  \wtT_j(z) \wtT_k(z) \dd \wtF(z) \,=\, \delta_{jk},~~\text{for all $j,k$}. \]
where, by a slight abuse of notation, $\wtT_j(z)$ denotes $T_j(z;\wtF)$. This orthonormality feature will be very useful for obtaining a neat computational algorithm that estimates the relevance function $d_x(z)$. 
\begin{rem}
Note the our empirical LP-polynomials $\wtT_j(z)$ are functions of ranks, since $\wtF(z_i)$ is equal to rank of $z_i$ divided by the sample size $N$. For that reason, we call them `LP-polynomials', where the the letter \textsc{L} denotes it is rank-based and \textsc{P} stands for polynomial.
\end{rem}

\subsection{Estimation of Relevance Function} \label{sec:estrel} 
The relevance function admits the following LP-orthogonal series representation:
\beq \label{eq:dexp}
d_x(z) \,:=\, d(F(z);Z,Z|X=x)\,=\,1\,+\,\sum_{j} \LP_{j|x} T_j(z;F).~~
\eeq
The goal is to estimate the unknown orthogonal LP-Fourier coefficients $\LP_{j|x}$, which determine the shape of the relevance function. To get a compact expression for these parameters, first note that
\beq \label{eq:LPcoef1}
\LP_{j|x} \,=\,\int_z d_x(z) T_j(z;F) \dd F(z).
\eeq
since $T_j$'s are orthonormal with respect to probability distribution $F$. Next, recall that $d_x(z)$ is actually $f(z|x)/f(z)$, by virtue of \eqref{eq:cdu}. Substituting this into \eqref{eq:LPcoef1}, we immediately get the following important result.
\begin{thm} \label{thm1}
The LP-Fourier coefficients $\LP_{j|x}$ admit the following conditional mean representation:
\beq \label{eq:LPcmean}
\LP_{j|x}\,=\,\int_z T_j(z;F)  f(z|x) \dd z\,=\,\Ex\big[T_j(Z;F)|X=x\big].~~
\eeq
\end{thm}
\vskip.4em
\begin{rem}
The practical significance of Theorem \ref{thm1} resides in the fact that we can now use the whole machinery of nonparametric and machine learning regression techniques to estimate the unknown LP-coefficients by regressing $T_j(z;F)$ on the multivariate feature $X$. Another important point to emphasize is that Theorem \ref{thm1} allows us to estimate the point-wise relevance function $d_x(z)$ by borrowing strength from the ensemble.
\end{rem}
\vspace{-.3em}
{\bf Robust LP-regression}. However, instead of directly regressing $T_j(z;F)$ on the covariates $X$, we approach it in a slightly matured way to inject robustness in the procedure\footnote[2]{
Robustness is critically important for the problem of reproducible inference from noisy heterogeneous data; see Sec. \ref{sec:repro} for more discussion.}. Our theory of estimation starts with the following important observation:
\begin{thm} \label{thm2} For mixed (discrete or continuous) $X$ we have the following important result: 
\beq \Ex[T_j(Z;F_Z)|X] = \Ex[T_j(Z;F_Z)|F_X(X)],~~ \text{with probability $1$}.
\eeq
\end{thm} 
This holds, owing to the fundamental fact of the quantile function: For a general (discrete or continuous) random variable we have $X=Q_X(F_X(X))$ with probability one \citep{parzen79}. The practical consequence of this result is that we can now approximate $\Ex[T_j(Z;F_Z)|X=x]$ by projecting onto the span of LP-bases $\{T_k(x;F_X)\}_{k\ge 1}$ of $X$.

\vskip.8em
{\bf LP-regression: Computational algorithm}.
From a computational standpoint, it amounts to simply running linear regression of $\wtT_j(z)$ on the LP-basis functions of $X$, which takes just one line in \texttt{R} by calling the \[\texttt{ML}\big(\wtT_j(z) \sim \wtT_{X}\big), ~~j=1,2,\ldots,m.\] 
where $\texttt{ML}$ can user-specified regression routine---e.g., simple linear regression (\texttt{lm}), lasso regression (\texttt{glmnet}), k-nearest neighbor (\texttt{knn}) regression, etc. Since our  style of nonparametric regression allows for easy integration with stepwise variable selection or other penalized methods (e.g., AIC, BIC, or even LASSO), it produces smooth nonlinear regression function with inbuilt robustness. A non-parametric bootstrap method to quantify the uncertainty of the estimated relevance function $\whd_x$ is discussed in Supplement H.
\vskip.8em

\begin{figure}
    \centering
    \includegraphics[width=\linewidth,keepaspectratio=true,trim=1.25cm 2.4cm 1.25cm 2cm]{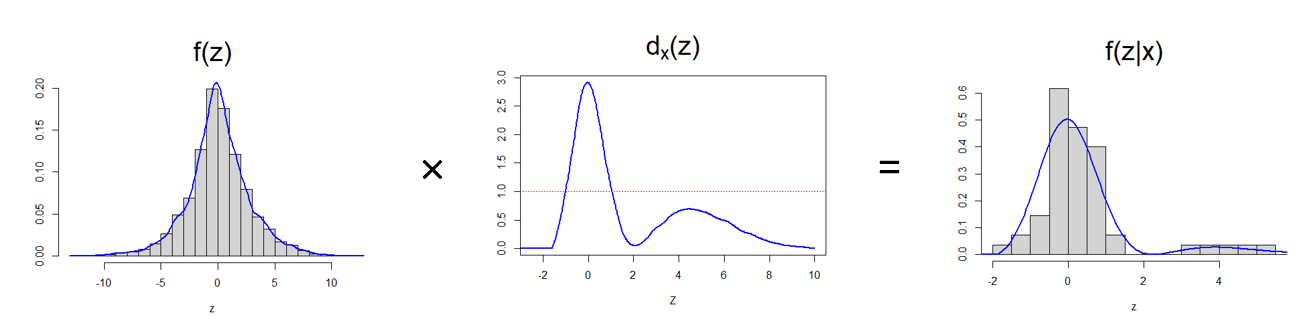}
    \vskip1.88em
    \caption{It shows the mechanics of global-to-local modeling for \texttt{funnel} data. The estimated relevance function at $x=30$ is shown in the middle plot. The conditional density estimate, shown in the last panel, seems to be `data-consistent' in the sense that it fits the observed data excellently.} \label{fig:simu_compden}
    \vspace{-.25em}
\end{figure}

{\bf Back to Funnel Example}. Figure \ref{fig:simu_compden} shows the mechanics of global-to-local modeling for the \texttt{funnel} data. First thing to note is that the estimated relevance function $\whd(z;Z,Z|X=30)$ deviates from uniformity, which means the conventional large-scale global analysis is not appropriate for cases with $x=30$. It also reveals how the distribution of the local z-values $\bbZ_x$ differ from the aggregated z-values $\bbZ$. In other words, much information supplied by the full ensemble is \textit{irrelevant} for the cases with $x=30$. We need some way of distilling the relevant information by refining m{\'e}lange of heterogeneous cases. This is done by our $d$-perturbative scheme (Eq. \ref{eq:g2l}), as demonstrated in Fig. \ref{fig:simu_compden}. The relevance function $d_x(z)$ modulates the global marginal $f(z)$ to yield the local $f(z|x)$ ---which has reduced the variability and an additional bump at the right tail, where those five true signals reside.

\subsection{Relevance Equitability Index}\label{sec:mrel} 
For a given target case with $X=x$, should we perform a combined or a customized analysis?
To satisfactorily answer this question we have to define a `relevance equitability index' (\texttt{REI}) to measure the degree relevance between the cases with $X=x$ and the full data. Interestingly, this information can be extracted from the shape of the relevance function $d_{x}(z)$. If the estimated relevance function is ``flat'' then it indicates that all the observations are equally relevant for the case in  hand--i.e., ``uniformity of relevance'' is a valid assumption. In that scenario, one can safely go with the usual global inferential methods. However, deviation from uniformity (as in Fig. \ref{fig:simu_compden}) suggests that customization is needed. In particular, one can define an appropriate \texttt{REI} by measuring the the unevenness of $d_x$:
\beq \label{eq:cust}
\CUST(x)\,=\,\int_0^1 \big\{d_x(u)-1\big\}^2 \dd u\,=\,\sum_{j} \big |\LP_{j|x}\big|^2.
\eeq 
A higher value of CUST-statistic indicates weaker relevance (i.e., higher degree of customization required) between the cases with feature $X=x$ and the overall sample.
\begin{rem}
The CUST-statistic can also be interpreted as a \textit{fairness-index}, which says how much it is fair to compare a given target case with the full ensembles of cases. Relevance and fairness are the two interrelated principles that underpin modern-day statistical inference. This is especially important for high-stakes decision making in applications such as health care, finance, hiring, criminal justice, etc.
\end{rem}
\vspace{-.4em}
Interestingly, one can even go to the extent of calculating the number of effective relevant samples available at each $x$: 
\beq \label{eq:rel} 
N_{{\rm rel}}(x) =N \times \rel(x),\eeq 
where
\[
\rel(x) = \dfrac{1}{1+\sum_j|\LP_{j|x}|^2} = \dfrac{1}{1+\CUST(x)}.
\]
\vskip.4em
When $d_x(u) \equiv 1$, i.e, all LP-Fourier coefficients are zero, we have $N_{{\rm rel}}(x)=N$, otherwise the effective sample size gets \textit{dampened} by the factor $\rel(x)$; see Supplementary C.
\begin{rem}
In summary, $\whd_x$ serves three purposes in customization: (1) quantification (measure of comparability), (2) characterization (nature of individualization required), and (3) synthesis of relevant samples. The last point is discussed in the next section.
\end{rem}

\subsection{Synthesizing LASER} \label{sec:sampler}
The relevance function provides an easy way to generate samples from the conditional distribution of $Z$ given $X=x$. The key is to use these samples as synthetic relevant cases  that permit one to ``zoom in'' on a specific target case. We call these simulated cases LASERs--they are specially-designed Artificial RELevant Samples. 

\vskip.64em
{\bf Learning whom to learn from}. Next we provide the algorithm to generate targeted LASER samples from the full aggregated data $z_1,\ldots,z_N$. Our global-to-local representation \eqref{eq:g2l} allows us to perform accept-reject-style sampling through $\whd_x$ to generate LASERs.
\vskip1em
~~~~~~~~~~~~~~~Algorithm 1. ~~Relevance Sampler:~Construction of \texttt{LASER}($N;x$)
\vspace{-1.24em}

\rule{\textwidth}{.8pt}
\vskip.1em  
\texttt{Step 0.} Input: The global $\bbZ = \{z_1,\ldots,z_N\}$; Target $X=x_0$, and $\whd_{x_0}(u)$.
\vskip.1em
\texttt{Step 1.} If the estimated $\whd_x(u)$ is ``flat'' uniform density (i.e., no customization warranted), then return the full data $\{z_1,\ldots,z_N\}$, else perform steps 2-5.
\vskip.1em
\texttt{Step 2.} Sample $z'$ from the global empirical cdf $\wtF$; In \texttt{R} perform: \[z' ~\leftarrow~ \texttt{sample}(z_1,\ldots,z_N,\,\texttt{size}=1,\, \texttt{replace}={\rm TRUE})~~~~\]
\texttt{Step 3.} Define $u'= \wtF(z')$. Generate $U \sim {\rm Uniform}(0,1)$.
\vskip.1em
\texttt{Step 4.}  Accept and set $z^* = z'$ if
\begin{equation*}
\whd_{x_0}(u') ~ > ~U\, \max_u\{\whd_{x_0}(u) \}
\end{equation*}
otherwise, discard $z'$ and return to Step 1.
\vskip.1em
\texttt{Step 5.} Repeat until we have obtained $N$ samples $\{z_1^*,z_2^*,\cdots,z_N^*\}$. We denote them by \texttt{LASER}($N;x_0$), which are samples from conditional distribution $\hf(z|x_0)$.\\
\rule{\textwidth}{.8pt}
\vskip1em
\begin{rem}
LASERs can alternatively be viewed as samples from the \textit{relevance-weighted} $z$-population (see step 2 and 4 above). The covariate-adaptive relevance weights $\whd_x(z_i)$ for $i=1,\ldots,N$ act as a \emph{``data sharpening''} tool to create tailor-made LASERs from the big  messy dataset. 
\end{rem}
\vspace{-.64em}



\section{Customized Statistical Inference} \label{sec:ci}
To deal with big messy datasets, the principle of relevance has to be an integral part of the laws of inference---which means we have to switch our attention from a one-size-fits-all global scheme to a more tailor-made one that takes into account the individual characteristics of the target case.  The question becomes, how to individualize a global inference method? Here we describe the principles and protocols of LASER-guided customized inference to answer this key question. The core idea is extremely simple: feed LASERs into your favorite global inference model to make it contextually adapted. 

\vskip1em
~~~~~~~~Algorithm 2. ~~LASER-guided customized inference: Algorithm in Pseudo-code
\vspace{-1.24em}

\rule{\textwidth}{.8pt}
\vskip.1em  
\texttt{Step 1.} Given $\{(x_i,z_i)\}_{i=1}^N$ the goal is to perform inference for cases with $X=x_0$.
\vskip.1em
\texttt{Step 2.} Generate \texttt{LASER}($N;x_0$) using the recipe of Algorithm 1, given in section \ref{sec:sampler}.
\vskip.1em
\texttt{Step 3.} Perform inference at $x_0$ by plugging-in lasers into global algorithm:
\[\texttt{global}\big(x_0\,;\, \texttt{LASER}(N;x_0)  \big).\]
Instead of \texttt{LASER}($N;x_0$), classical global methods use $\bbZ =\{z_1,\ldots,z_N\}$ as the fixed comparison set for \textit{all} cases. This section, through several examples, attempts to demonstrate that learning an appropriate relevant comparison set is often a prerequisite for a valid large-scale inference method, especially when we are dealing with dissimilar cases.
\\
\rule{\textwidth}{.8pt}
\vskip.55em
The most attractive part of this algorithm is its simplicity and generalizability--and the most crucial part of this algorithm is to properly synthesize the lasers, which we feed into (any user-preferred) global inference machine. The following histograms show the \texttt{LASER}($N;x$) for the funnel example at $x=30$ and $60$.
  \begin{figure}[ht]
  \vskip1em
        \centering
\includegraphics[width=.44\linewidth,trim=1cm 1.4cm 1cm 0cm]{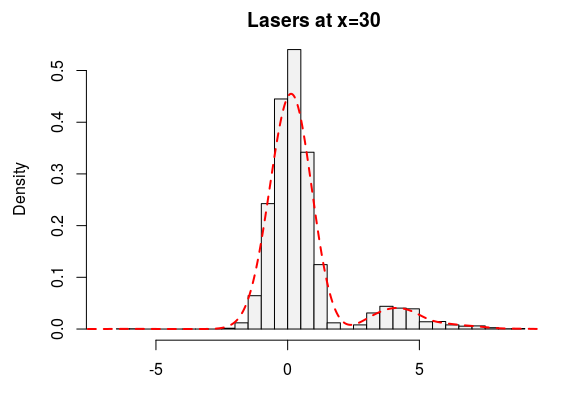}~~~~~~~~~
 \includegraphics[width=.44\linewidth,trim=1cm 1.4cm 1cm 0cm]{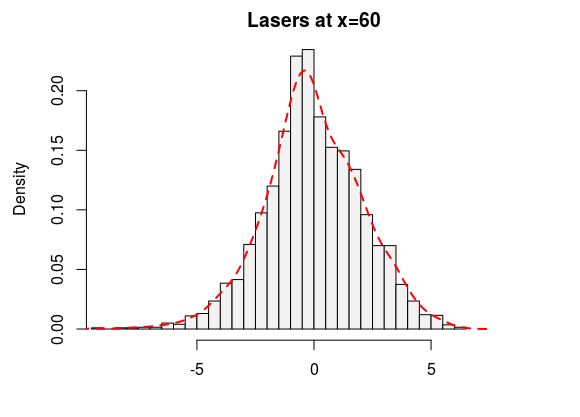}
 \vskip.35em
\caption{Histograms of the lasers at $x=30$ and $x=60$ for the funnel data. Notice the contrasting shapes, in particular, the difference in the width of the two histograms.} \label{fig:lasers}
\vspace{-.15em}
\end{figure}
As seen in Fig. \ref{fig:lasers}, lasers capture the ``natural variation'' that is present at $x=30$ and $60$.  For example, $z=4$ can be considered as ``large'' for the population with $x=30$, but is a typical occurrence for $x=60$ population. The same argument holds for the effect-size estimation problem: $z=4$ should be shrinked (towards zero) much more aggressively for $x=60$ cases, compare to $x=30$. The bottom line is context matters. Lasers allow us to contextualize any global inference method,  in one step.
\subsection{MicroInference} \label{sec:micro}
Given the z-values $z_1,\ldots,z_N$, the local-false discovery rate \citep{efron2010book} is defined as
\beq \label{eq:glocfdr}
\fdr(z)=\Pr(\text{null}|Z=z)= \dfrac{\pi_0 f_0(z)}{f(z)}, 
\eeq
where the last equality follows from the Bayes rule, $\pi_0=\Pr(\text{null})$, $f_0(z)$ is the null density, and $f(z)$ denotes the marginal distribution of all the $z$'s. We seek to develop a computable theoretical framework for covariate-adaptive false-discovery method.
The following theorem shed light into the practical challenges that need to be dealt with before developing a mathematically precise solution.


\begin{thm} \label{thm:fdr}
The conditional false discovery rate (fdr) function $\fdr(z|x)$ admits the following `global-to-local' representation (close in spirit to equation \ref{eq:g2l}) in terms of the marginal $\fdr(z)$:
\beq \label{eq:fdrexp} 
\fdr(z|x)~=~\fdr(z)\left[\frac{\pi_0(x)}{\pi_0} \times \frac{f_0(z|x)}{f_0(z)} \times  \frac{1}{d(F_Z(z);Z,Z|X=x)} \right],\eeq
where $f_0(z|x)$ is the null distribution of $Z|X=x$, and $\pi_0(x)$ is $\Pr({\rm null}| X=x)$. 
\end{thm}
{\bf Proof}. The trick lies in expressing the conditional fdr function as
\beq \fdr(z|x)\,=\,\Pr(\text{null}|Z=z,X=x)\,=\, \dfrac{\pi_0(x) f_0(z|x)}{f(z|x)}, \eeq
which by virtue of eq. \eqref{eq:glocfdr} can be re-written as
\[ \fdr(z)\left[\frac{\pi_0(x)}{\pi_0} \times \frac{f_0(z|x)}{f_0(z)} \times  \frac{f(z)}{f(z|x)} \right].  \]
Finally, apply eq. \eqref{eq:g2l} and substitute $1/d_x(z)$ for the ratio $f(z)/f(z|x)$ to finish the proof.
\vskip.4em

The derived theory is undoubtedly beautiful but it contains uncalculable parameters! Let's focus on the ``relevance correction'' part inside the 
square brackets of \eqref{eq:fdrexp}: (i) The first factor $\pi_0(x)/\pi_0 \approx 1$ for most practical problems; (ii) the last factor is the ``well-behaved'' $d_x(z)$ function, whose estimation is already discussed in Section \ref{sec:estrel}. (iii) Finally, we are left with the factor in the middle: ratio of relevant null $f_0(z|x)$ to the global null $f_0(z)$. How to empirically estimate the relevant null? Efficiently estimating the parameters of $f_0(z|x)$ is difficult (if not impossible), as we have too little direct data available at $X=x$. The current literature bypasses this problem by \textit{assuming} that $X$ is independent of $Z$ under the null hypothesis, i.e., $f_0(z|x)=f_0(z)$. But is this a sensible assumption?

  \begin{figure}
        \centering
        \begin{subfigure}{.466\textwidth}
        \caption{}
            \includegraphics[width=\linewidth,trim=1cm 1cm 0cm 0cm]{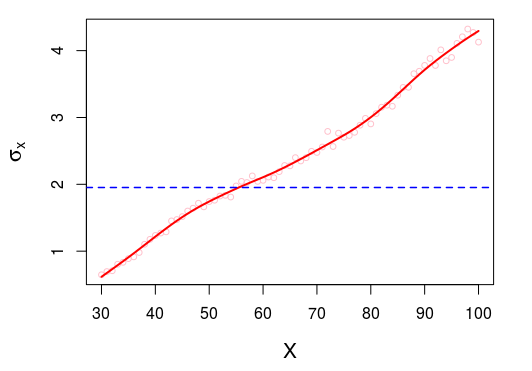}
        \end{subfigure}~~~
        \begin{subfigure}{.466\textwidth}
        \caption{}
            \includegraphics[width=\linewidth, trim=0cm 1cm 1cm 0cm]{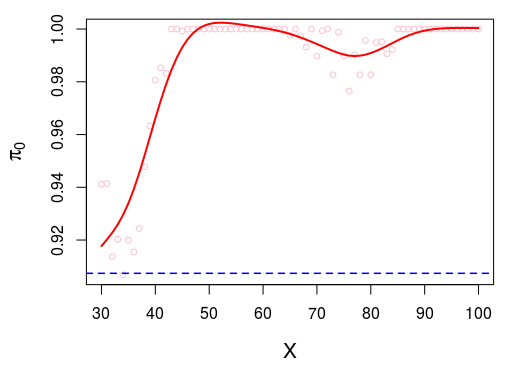}
        \end{subfigure} 
        \vskip1.6em
        \caption{Funnel data: How do the parameters of the relevant empirical nulls $\cN(\mu_0(x), \si_0(x))$ and $\pi_0(x)$ change with $x$? As the data is already centered (i.e., $\mu_0(x)=0$, for all $x$), we only focus on the $ \si_0(x)$ and $\pi_0(x)$ which is show respectively in the panel (a) and (b). They are calculated by applying global empirical null estimation algorithm (locfdr) on \texttt{LASER}($N;x$), at each $x$. The blue dotted lines denote the global empirical null parameter values in each plot.}  \label{fig:emnull}
    \end{figure}

 \begin{figure}
        \centering
        \includegraphics[width=.6\linewidth,trim=1cm 0cm 1cm 1cm]{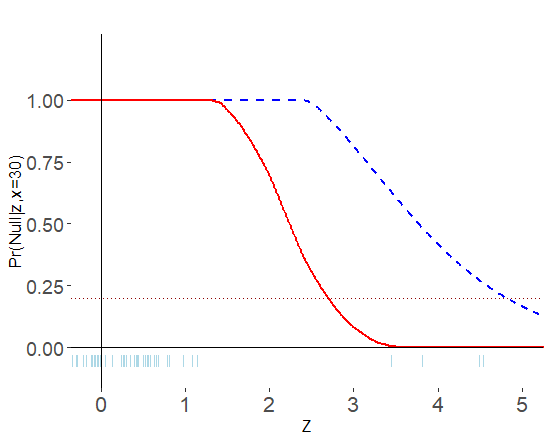}
        \caption{The shape of $\fdr(z|x=30)$ for the \texttt{funnel} example. The red is our customized fdr and blue is the global one. The horizontal red dotted line denotes $0.2$ threshold.}
           \label{fig:fdrs}
    \end{figure}

It is actually a dangerous assumption, especially when we are dealing with large heterogeneous data. Fig. \ref{fig:emnull} shows how the different parameters of the relevant null are changing as a function of $x$ for the \texttt{funnel} example. The most dramatic one, among these two, is the leftmost one, Fig. \ref{fig:emnull}(a),  which shows how different the standard deviations are between the global and the relevant null. This makes the ratio of these two nulls 
\[ f_0(z|x)/f_0(z) \,\approx\, \mathlarger{e^{-\frac{z^2}{2} (1/\si_0^2(x) - 1/\si_0^2)}},\] 
since the mean parameters are all practically zero. As seen from Fig. \ref{fig:emnull}(b), $\si_0(x) < \si_0$ for all $x$ less than $50$, which means the ratio of the nulls will exponentially decrease for large $|z|$. This explains why the estimated conditional fdr function $\fdr(z|x=30)$ in Fig. \ref{fig:fdrs} sharply bends inward and successfully detects all the true signals at the extreme right.
\begin{rem}
It is our view that insufficient regard for the relevant (empirical) null $ f_0(z|x)$ is the root cause why current large-scale inference methods fail so miserably. Supp. Appendix D reviews this issue of estimating `relevant empirical null' in more details.
\end{rem}
\begin{rem}
Theorem \ref{thm:fdr} provides an indirect two-step estimation recipe for $\fdr(z|x)$ by going through the marginal $\fdr(z)$ function. In practice, we can do a quick approximation in a much simpler and more direct manner by simply feeding point-wise \texttt{LASER}($N;x$) into the global inference engine, as implemented in Fig. \ref{fig:fdrs}. In that sense, $\fdr(z|x)$ can be considered as a ``synthetic model''.\footnote[2]{A side remark: this idea of `synthetic' modeling might turn out to be useful to build `privacy-preserving' statistical inference machines at an individual level.} See Supplementary G for a step-by-step description of our microinference procedure. 
\end{rem}

\begin{rem}
There is an alternative way to express $\fdr(z|x)$ by conditioning on $z$ (instead of $x$ as done in Theorem \ref{thm:fdr}):
\beq \label{eq:lsieq}
\Pr(\text{null}|Z=z,X=x)\,=\, \dfrac{\pi_0 f_0(z)}{f_Z(z)} \times \dfrac{f_0(x|z)}{f(x|z)}.~~~
\eeq
The first factor is simply the marginal $\fdr(z)$. For $X$ \textit{discrete}, the second factor can be written as a conditional probability of $X=x$ given $Z$:
\[\Pr(\text{null}|Z=z,X=x)= \fdr(z) \cdot \dfrac{\Pr_{{\rm null}}(X=x|Z=z)}{\Pr(X=x|Z=z)} = \fdr(z) \cdot \dfrac{\varpi^0(x|z)}{\varpi(x|z)},~~\]
which matches with Eq. (2.16) of \cite{efron2008com}, after substituting $\varpi^0(x|z)/\varpi(x|z)$ by $R_x(z)$; also compare with Theorem 10.3 of \cite{efron2010book}. Note that we prefer conditioning by $x$ to retain the applicability of our formula for mixed (either discrete or continuous) multivariate $X$.
\end{rem}
\vspace{-.5em}





\subsection{MacroInference} \label{sec:macro}
Our goal is to perform a full-scale search and combing operation to locate the hidden signals, by retaining the individuality of each case.

\begin{figure}
        \centering
        \includegraphics[width=.45\linewidth,trim=1cm 0cm 1cm 1cm]{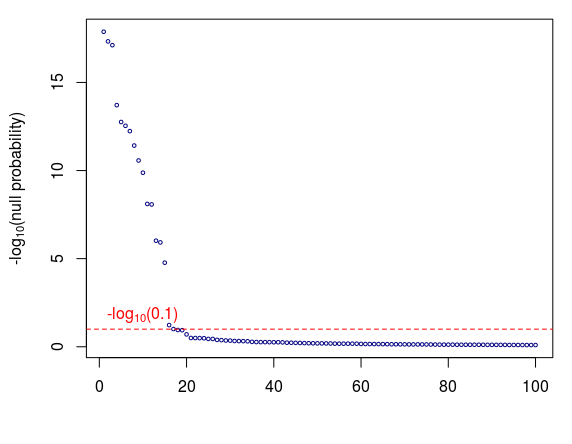}~~
        \includegraphics[width=.5\linewidth,trim=0cm 0cm 0cm 0cm]{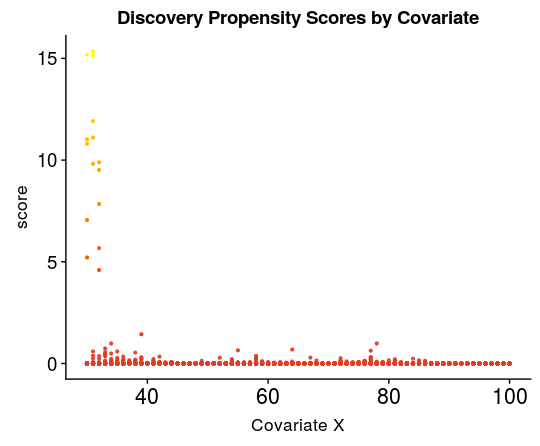}
        \caption{(color online) The sorted DPS-score of the top 100 cases for the \texttt{funnel} example. Left plot: The ``gap'' in this sorted list is informative, which automatically finds the `batches' (+ \textit{in which order} to investigate them) worthy of careful followup study. The red dotted line represents the locfdr threshold ($\approx 2 \al$) in the $-\log_{10}$ scale for $\al=0.05$. The right panel shows the scatter ($x_i, {\rm DPS}_i$), for $i=1,.\ldots,N$. This diagnostic plot can be used to detect the signal-prone regions (here, as clearly visible $x=30,31$ and $32$).}
         \label{fig:dps}
    \end{figure}

 \begin{figure}
        \centering
        \includegraphics[width=.48\linewidth,trim=2cm 0cm 0cm 1cm]{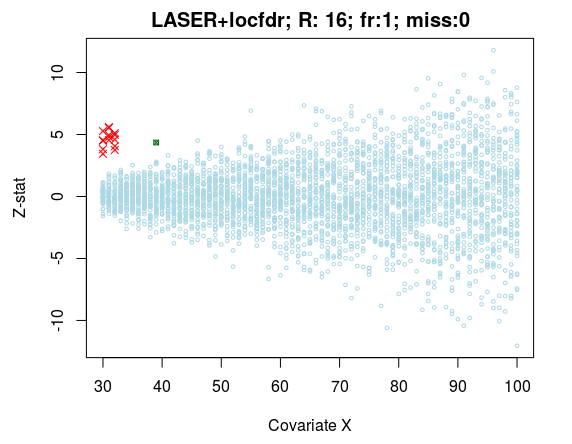}
          \includegraphics[width=.48\linewidth,trim=0cm 0cm 2cm 1cm]{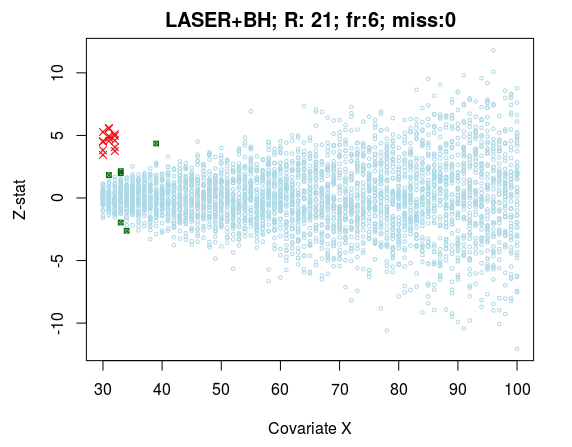}
       \caption{Output of LASER-guided customized multiple testing result at level $\al=0.05$, using local false discovery rate and Benjamini–Hochberg as the global procedures. `R' stands for number of rejections, `fr' means number of falsely declared signals, and `miss' denotes number of true signals missed.}  \label{fig:comp}
\end{figure} 





{\bf Stage 1: Triage}. This is a process of prioritizing or sorting cases based on their discovery proneness. For case$_i$, define `Discovery Propensity Score'
\beq {\rm DPS}_i~=~ -\log_{10}\big\{\Pr({\rm null}|z_i,x_i) \,\big\} = -\log_{10} \big\{\fdr(z_i|x_i)\,\big\},~~ {\rm for}~ i=1,\ldots,N.
\eeq
DPS-values act as an index to rank the cases. Fig. \ref{fig:dps} shows the DPS-plot for the \texttt{funnel} data, which correctly separates (notice the ``gap'') the $15$ signals from the rest of the null cases. Investigators can use this ordered list of cases for more detailed follow-up studies.
\begin{rem}
Unlike p-values, the DPS-values can be used as summaries of statistical evidence, since they provide a direct assessment of the
probability that a finding is spurious.
\end{rem}

\vspace{-.45em}
{\bf Stage 2: Select}. Here we ``select'' a small number of the most promising results by applying the false discovery threshold. Fig. \ref{fig:comp} shows the remarkable performance of the LASER-guided multiple testing procedures; contrast this with Fig. \ref{fig:global_simu}. We have used both local false discovery (locfdr) and Benjamini Hochberg (BH) as the choice of global large-scale testing methods. \nocite{BH95}


\vspace{-.5em}
\subsection{DTI Neuroscience Data Analysis} \label{sec:app1}

\begin{figure}[t]
\vspace{-1em}
    \centering
        \includegraphics[width=.57\linewidth,trim=1cm 1cm 1cm 1cm]{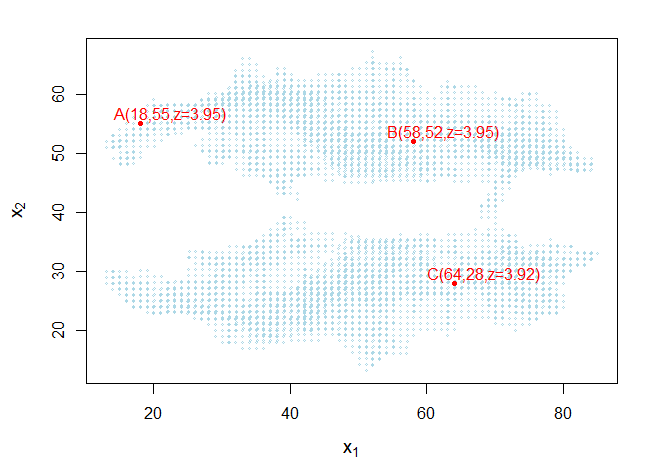}
    \vskip.5em
    \caption{The 2D map of DTI data. Three target cases (A,B, and C) with similar z-values are marked with red dots.} 
    \label{fig:dti_overview}
\end{figure}



Here we apply our customized large-scale testing procedure to a data set obtained from diffusion tensor imaging (DTI). This study compares brain activity of $n_1=6$ dyslexic children with $n_2=6$ normal controls. We are given two-sample z-values $z_1,\ldots,z_N$ of $N=15,443$ voxels, along with the location information: $X_{1i}$ (distance from the back of the brain) and $X_{2i}$ (distance from the right of the brain). Our primary interest is in: (i) microinference: to estimate the customized local fdr curve for voxels A,B, and C; and (ii) macroinference: to locate the significant voxels; see  Fig. \ref{fig:dti_overview}.

\begin{figure}[t]
    \centering
    \begin{subfigure}{.32\textwidth}
        \caption{Case A}
            \includegraphics[width=\linewidth]{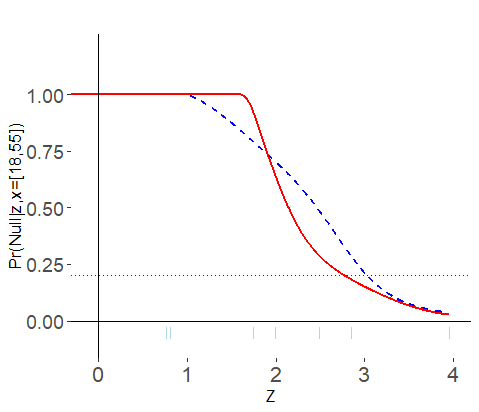}
        \end{subfigure}
        \begin{subfigure}{.32\textwidth}
        \caption{Case B}
            \includegraphics[width=\linewidth]{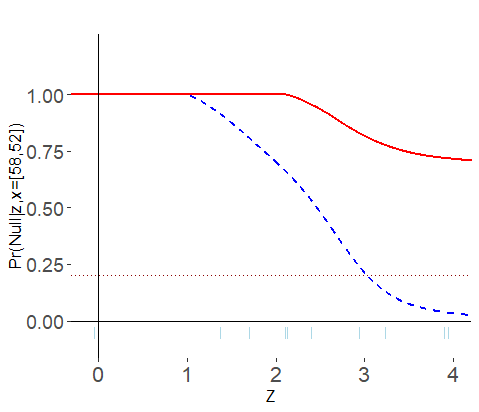}
        \end{subfigure}
        \begin{subfigure}{.32\textwidth}
        \caption{Case C}
            \includegraphics[width=\linewidth]{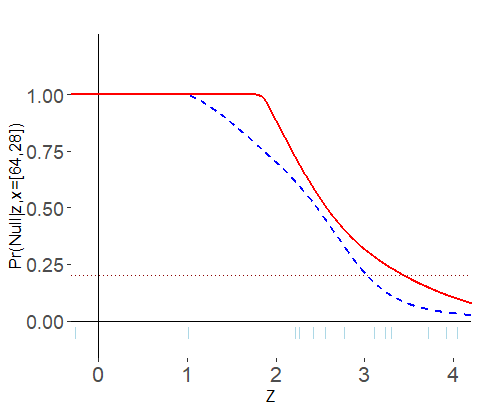}
        \end{subfigure}
        \includegraphics[width=.4\textwidth]{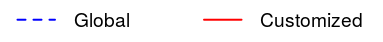}
    \caption{The fixed global fdr function is denoted by the blue dotted line. Our customized fdr curves for voxels A, B and C are shown in red. Note their individually adaptive shapes. All the computations are done using the \texttt{R}-package \texttt{LPRelevance}.}
    \vspace{-.5em}
    \label{fig:dti_fdr}
\end{figure}

{\bf Microinference}. The blue dotted curve in Fig. \ref{fig:dti_fdr} denotes the full-data based (global) fdr function. We are interested in individualizing this global fdr function at a voxel-level. Let's start with the case A, which has the z-value $3.95$ at the location $(x_{1A}=20, x_{2A}=55)$. At exactly the same location, we have only nine other relevant voxels! This minuscule size makes it impossible to estimate the customized-fdr function from direct relevant samples (this weird phenomenon was illustrated in Fig. \ref{fig:relpara}). To tackle this problem, we generate \texttt{LASER}($N;x_{1A},x_{2A}$) and plug them into the locfdr function to generate the red curve in Fig. \ref{fig:dti_fdr} (a). We follow the same procedure for the other two cases, B and C. What's most interesting about these plots is the contextually-adaptive \textit{shape} of the $\fdr(z|x)$ functions, even when the z-values are almost same!

\begin{figure}[ht]
    \centering
    \centering
    \includegraphics[width=.48\linewidth,trim=1cm .25cm .5cm .5cm]{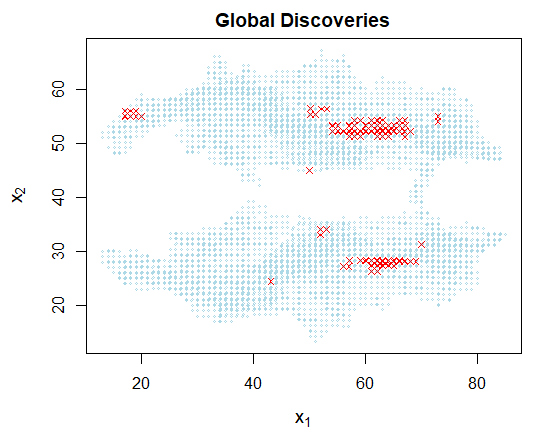}~~~~
\includegraphics[width=.48\linewidth,trim=1cm .25cm .5cm .5cm]{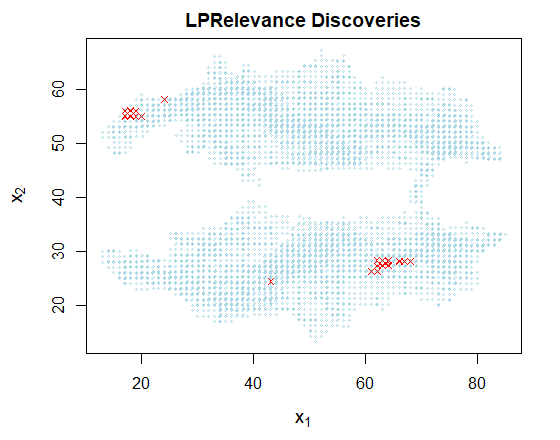}\\[1.65em]
\includegraphics[width=.48\linewidth,trim=1cm .25cm .5cm .5cm]{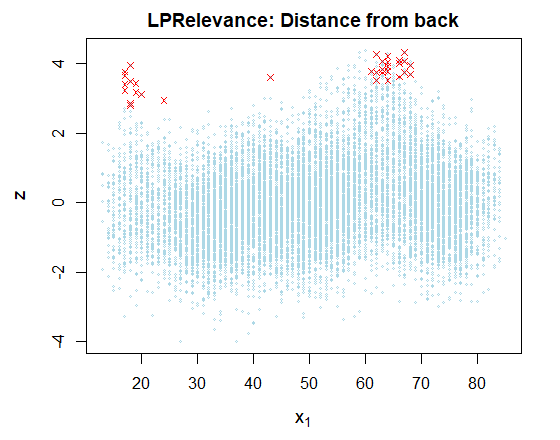}~~~~
\includegraphics[width=.48\linewidth,trim=1cm .25cm .5cm .5cm]{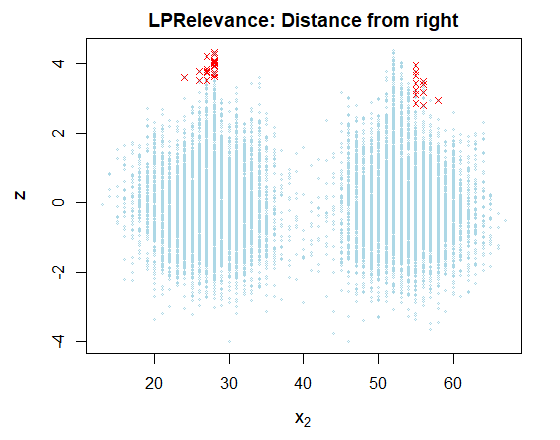}
\vskip1em
    \caption{Top row: Comparison of macro inference results using global and our LASER-guided customized locfdr. Bottom row: Displays our findings separately as a function of $x_1$ and $x_2$ for easy interpretation and visualization. The red crosses are `significant' voxels.}
    \label{fig:dti_macro}
    \vspace{-.5em}
\end{figure}

{\bf Macroinference}. Next, we move to the question of macroinference: can we find a few interesting, differentially expressed voxels? Our result is summarized in Fig. \ref{fig:dti_macro}. Global locfdr method compares each voxel with all the remaining $N-1$ voxels for assessing significance, which is clearly inadvisable due to heterogeneity. If we define `signal' as the exceptional cases among their own tribe, it makes complete sense to compare each voxel with its own specially-designed LASERs for a fair comparison. Our customized locfdr declares $33$ voxels to be significant at $\al=0.1$. The global version, meanwhile, finds $190$ discoveries. Supplementary Fig. \ref{fig:suppDTI} takes a closer look at the cluster of $111$ voxels around the left frontal area of the brain, which were declared significant by the global locfdr but avoided tactfully by the customized one. Surprisingly, all of these additional cases clump together at the top of the heterogeneity wave, near $x_1=60$ and $x_2=55$. This makes us suspect that they look ``big'' because of the unaccounted heterogeneity. The main point here is: raw magnitude of a case does not matter; what matters is how big a specific case is \textit{with respect to} its own LASERs. It's all relative! 

\begin{figure}
            \centering
            \vspace{-2.5em}
    \includegraphics[width=.645\linewidth,trim=2cm 0cm 3cm 0cm]{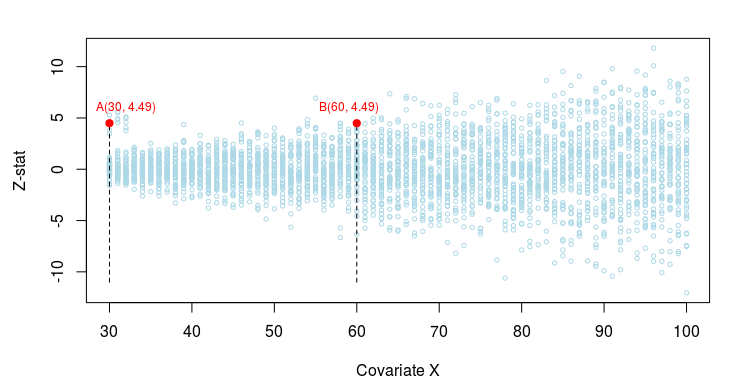}~~~~~~~~~~~~~~~~\\[1.7em]
    \includegraphics[width=.84\linewidth,trim=1cm 0cm 1cm .5cm]{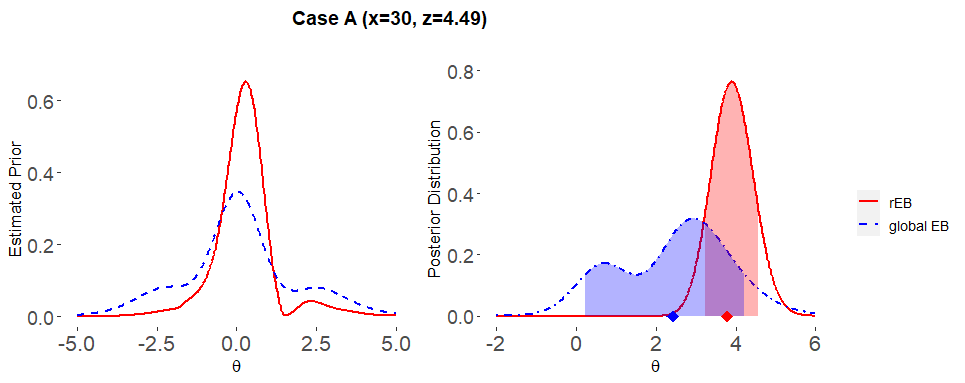}\\[.34em]
    \includegraphics[width=.84\linewidth,trim=1cm 0cm 1cm .5cm]{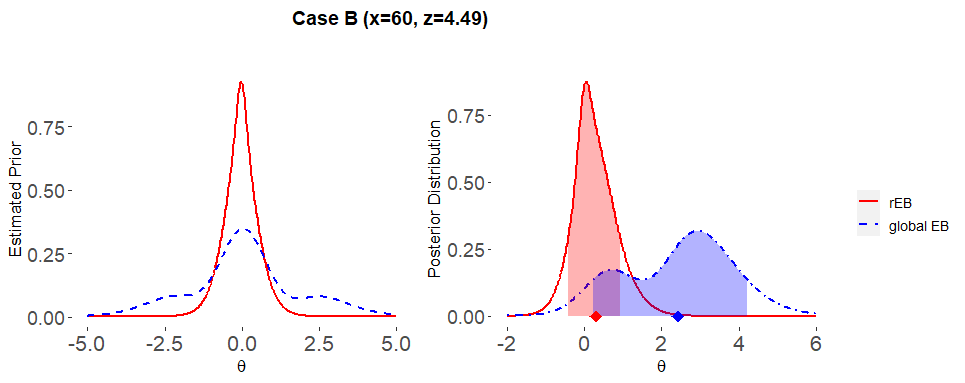}
    \vspace{-.525em}
    \caption{Estimated prior and posteriors for two cases, A and B. Both have identical z-value, yet one is signal and the other is noise. The global empirical Bayes (gEB) prior and posterior are denoted by blue dotted curves, which are unchanging. The red curves are rEB results, changing with the contextual variable; contrast the sharpness of the rEB priors around zero. The gEB estimate (mean): $2.42$, denoted as blue dot. The rEB estimates (mean): for case A (true signal) is $3.78$ and for case B (noise) $0.29$ are denoted as red dots. The shaded areas denote 80\% highest posterior density (HPD) intervals. The rEB produces much more precise (properly centered narrower HPD) estimates---addressing both the selection and relevance bias. For a step-by-step guide on the rEB algorithm, see Supplementary F.}
            \label{fig:EB} 
        \end{figure}

\subsection{Empirical Bayes Inference} \label{sec:emb}
We now shift our focus from testing to estimation. Given a large number of sample $z$-statistic $z_i \sim \cN(\te_i,\si_i^2)$, the goal is to estimate the unknown mean parameters (also called effect sizes) $\te_i$, especially for the non-null cases. By now it is well-known that empirical Bayes provides a simple and elegant approach to effect-size estimation \citep{efron2011tweedie} by enabling one to `learn from the experience of others.' However, the basic premise of empirical Bayes relies on the assumption that we are given a bag of samples that are relevant to each other---which, of course, is questionable for most real-world practical problems. Stuck in such a predicament, how should we proceed? 
\vskip.44em
{\bf Global to Individualized Relevant Prior}. The core idea is remarkably simple: rather than lumping heterogeneous, unrelated cases all together, what if we used LASERs to estimate the context-aware ``personal'' prior? Fig. \ref{fig:EB} shows two cases: A ($x=30,z=4.49)$ and B ($x=60,z=4.49$). They have the same z-value but in two different contexts, captured by the covariate $x$. The global empirical Bayes prior\footnote[2]{estimated using the algorithm prescribed in \cite{deep18nature}, which is implemented in the R-package \texttt{BayesGOF} \citep{BayesgofR}.
 However, one can use any other method--e.g., Efron's deconvolution \citep{efron2016} or Koenker's NPMLE \citep{koenker2016}. The important point here is \textit{not} how to estimate an EB prior, but how to design `right' relevant samples to learn from.} is shown in the blue dotted line which, by design, does not change with $x$. On the contrary, LASER-guided empirical Bayes priors show interesting differences: $\pi_A(\te)$ has a longer tail with a slight bump around $2.5$, and $\pi_B(\te)$ has a much sharper peak around zero. It's impact is clearly visible on the posterior distributions.  Table \ref{tab:eb} summarizes the effect-size estimates for global as well as relevant empirical Bayes (rEB), which shows the adaptive shrinkage property of our rEB method.
   \begin{table}[ht]
        \centering
        \caption{Effect-size estimates (posterior mean) for cases A and B: Comparing global empirical Bayes (gEB) with contextually-tailored relevant empirical Bayes (rEB) analysis.} \label{tab:eb}
        \begin{tabularx}{.96\linewidth}{h f f}
        \toprule
            & $\hat{E}[\Theta|x,z]$  & $80\%$ HPD Interval  \\
        \midrule
            global-EB: & $  2.42 ~ $  & $(0.23,~ 4.20)$\\
            rEB:A &$ 3.78$ & $( 3.23,~ 4.58 )$\\
            rEB:B & $0.29$  & $(-0.41,~  0.93 )$\\
        \bottomrule
        \end{tabularx}
        \vspace{-.5em}
    \end{table} 

\begin{rem} A few things are evident from this study

~~1. Our rEB framework \textit{simultaneously} balances two kinds of errors: selection bias \citep{efron2011tweedie}, and relevance bias \citep{efron1972EB}. Without relevance-correction, the usual global EB analysis would produce faulty effect-size estimates. The reason being, \textit{where} to shrink and \textit{how much} to shrink is directly related to the shape for the relevant EB prior, which is often very different from the global EB prior, especially in the tails; see Fig. \ref{fig:EB}.
To the best of our knowledge, no previous research has achieved this dual goal of balancing the selection as well as relevance bias; rEB made it possible by selectively borrowing information from other \textit{relevant} cases through LASERs.

~~2. Redeeming the curse of a real winner:  From Table \ref{tab:eb}, note that the global EB estimate of $\hte_A$ is $2.42$ and the relevance-corrected estimate is $3.78$. Our rEB-adjustment prevents over-shrinkage---unfairly pulling $z_A=4.49$ towards the zero effect for cases with $x=x_A$. As a result, rEB produces a more \textit{fairer} effect-size estimate,  by taking context into account.

~~3. Our style of empirical Bayes analysis is completely data-driven, which avoids the appearance of arbitrariness resulting from guessing the different parametric forms of the relevance functions \cite[p.14]{efron2011tweedie}.

~~4. One other important point is that LASERs reduce the direct contact of the prior with the actually observed data, and hence alleviate the ``double-dipping'' problem.
\end{rem}
\vspace{-.34em}
{\bf Variance Reduction by Model-averaging}. For a more precise answer perform the following model averaging: (i) generate $B$ (say, $B=10$, which is often enough) bags of parametric bootstrapped \texttt{LASERs}($N;x$) from the estimated $\whd_x(z)$; (ii) for each bag, compute the locfdr curve and the empirical Bayes posterior distribution; (iii) finally, return the ``averaged'' estimated curve (averaging over $B$ runs). This ``bagged estimate'' reduces the variability  of a single LASER-based model without  affecting the bias. 
\vskip.25em
{\bf Connection with Regression-adjusted Empirical Bayes}. The regression-adjusted empirical-Bayes approach starts with the following model: $z_i \sim \cN(\mu_i,\si_0^2)$, and $\mu_i \sim \cN(\al + \be x_i,\tau^2)$. Surely, instead of simple linear regression, one can use any nonparametric method, but the basic idea is simple: take out the mean heterogeneity by fitting a curve
\beq \label{eq:ebreg} y_i = z_i - (\widehat{\al} + \widehat{\be} x_i),~i=1,\ldots,N\eeq
and then apply global empirical Bayes method on $y_i$'s to make a prediction for a specific case. This is a completely justified model, provided we assume the heterogeneity is affecting only the mean of $Z|X=x$, which we call the ``first-order'' covariate adjusted model.  Unfortunately, a practical statistician might find it an overly simplified and unrealistic assumption. Consider for example the \texttt{funnel} data, where the conditional mean is not even changing, but there exists substantial higher-order heterogeneity. Thus the real question is whether we can develop a general covariate-adjusted empirical Bayes framework that \textit{includes} the first-order regression-adjusted model as a \textit{special case}. Remarkably, the answer is yes, as illustrated in the next section.

\begin{figure}
    \centering
    \begin{subfigure}{.48\linewidth}
    \caption{Scatter Plot}
        \includegraphics[width=\linewidth,trim=1cm 0cm 0cm .5cm]{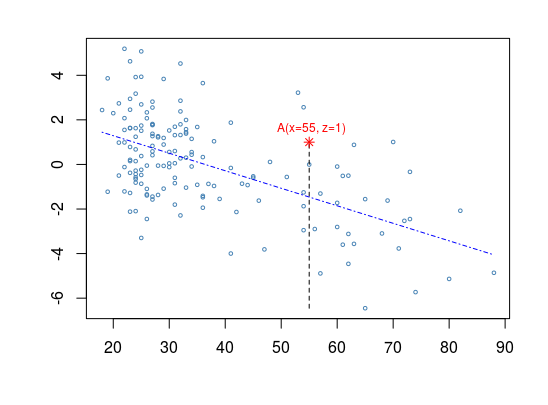}~~~~~
    \end{subfigure}
    \begin{subfigure}{.48\linewidth}
    \caption{Regression-Adjusted}
        \includegraphics[width=\linewidth,trim=1cm 0cm 0cm .5cm]{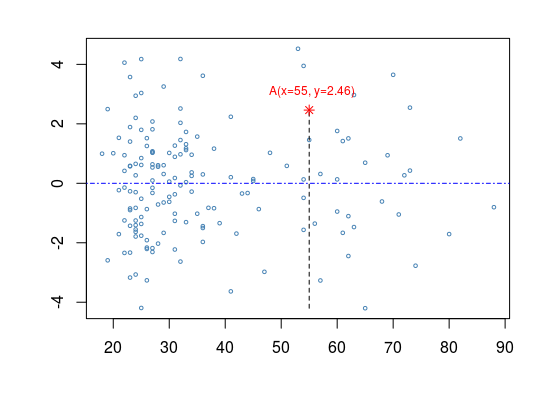}
    \end{subfigure}\\[1.5em]
    \begin{subfigure}{.478\linewidth}
    \caption{Relevance Function}
        \includegraphics[width=\linewidth,trim=1cm 0cm 0cm .5cm]{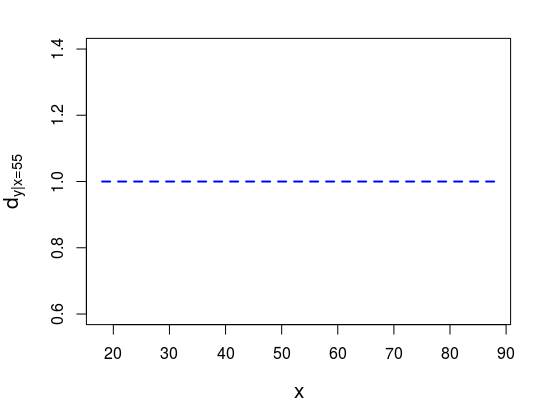}
    \end{subfigure}~~~~~~~
    \begin{subfigure}{.482\linewidth}
    \caption{Posterior}
         \includegraphics[width=\linewidth,trim=1cm 0cm 0cm .5cm]{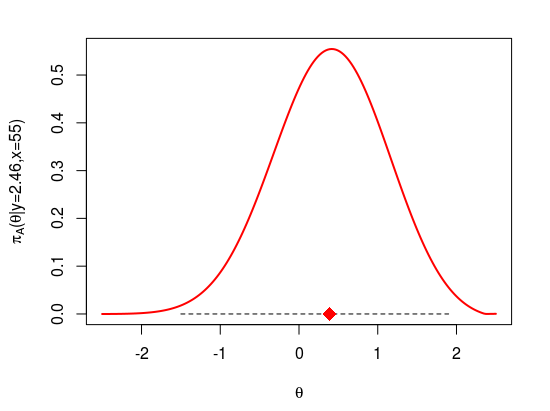}
    \end{subfigure}
    \vskip1em
    \caption{Steps of kidney data analysis: (a) scatter plot of age verses kidney function; The blue dotted line denotes the least-square regression function $2.86-0.0786x$; The red `*' is the target case A for which inference is sought. The regression estimate at $x=55$ is $2.86-0.0786 \times 55=-1.463$. (b) flattening step $z \rightarrow y$: the regression-adjusted $(x,y)$ plot, where $y_i=z_i - (2.86 - 0.078x_i)$; (c) The estimated relevance function $\whd_x(y)$, which interestingly takes the flat uniform shape; (d) The estimated posterior distribution  $\pi_{A}(\te|x=55,y=2.46)$ with posterior mean $0.385$.} \label{fig:kidney}
        \end{figure}

\subsection{Kidney Data  Analysis} \label{sec:app2}
Fig. \ref{fig:kidney} displays the age and kidney function of $N = 157$ volunteers. Higher scores indicate better function. We are interested in the following question \cite[Ch. 1.4] {efron2010book}: What is the empirical Bayes shrinkage estimate for the case A ($x=55,z=1$), denoted by the red `*' sign in Fig. \ref{fig:kidney} (a)? The main steps of our analysis are summarized below:


     \vskip.29em
\texttt{Step 1.} Flattening: Looking at the data, the first obvious thing to do is to take out the mean-heterogeneity by fitting a regression-smoother. The fitted least-square line is shown in Fig. \ref{fig:kidney} (a). Next, we construct $y_i$'s by subtracting $z_i$ from $\widehat{\al} + \widehat{\be} x_i$; see the panel (b). We call this process ``flattening'' of scatter.
     
     \vskip.29em
\texttt{Step 2.} Estimation of the Relevance Function: Can we make a better inference for the target case A (see panel B) by learning from the experience of other 156 volunteers? Yes, if they are comparable---i.e., if the $y$-values around $x=55$ have the same statistical (distributional) properties as the full data. This equatability information is stored in the relevance function $d_{x=55}(y)$, which contrasts the conditional density $f(y|x=55)$ with the marginal $f(y)$. We apply the theory of section  \ref{sec:estrel} to estimate the unknown coefficients of $d_x$:
\beq 
\hat d_x(y) \,=\, 1 + \sum_{j=1}^m \widehat \LP_{j|x}\, T_j(y;F_Y), ~~\text{at $x=55$}.~~~ 
\eeq
The BIC-selected coefficients all turn out to be zero: $\widehat{\LP}_{j|x=55}=0$ for $j=1,\ldots,m=6$. That means the relevance function is ``flat'': $\whd_{x=55}(y) = 1$; see Fig. \ref{fig:kidney}(c).
     \vskip.29em
\texttt{Step 3.} ``Uniformity of Relevance'' Test: The flat shape of the estimated relevance function indicates that there is no ``excess'' heterogeneity in $f(y|x=55)$ relative to the ensemble, and therefore the other $N=156$ cases can be taken as admissible comparison set for borrowing information. Relevance function acts as a formal nonparametric exploratory test to validate this assumption. Accordingly, the relevance sampler (see Section \ref{sec:sampler}) returns the full observed data $\{y_1,\ldots,y_{157}\}$ as \texttt{LASERs} for case $A$.
     \vskip.29em
\texttt{Step 4.} Posterior Analysis of A: We borrow strength from the $y$-ensemble, to perform microinference for donor A ($x=55,z=1$). Model: $y_i |\te_i \sim \cN(\te_i,\si_0^2)$ where we have used 
\[\si_0= \texttt{IQR}(y)/1.3489 =1.79, \]
\texttt{IQR} stands for interquartile range. Fig. \ref{fig:kidney}(d) shows the estimated (empirical  Bayes) posterior distribution $\pi_A(\te|y=2.46,x=55)$ with posterior mean $0.385$.
     \vskip.29em
\texttt{Step 5.} Empirical Bayes Correction: Finally, we transform the $y$-domain answer in the original $z$-domain: $0.3845 -1.46 =-1.0755$. Note that this ``corrects'' the frequentists' regression estimate $-1.46$ by a factor of $0.385$.
    \vskip.29em
This whole rEB analysis can be implemented in a few lines using \texttt{LPRelevance} R-package. See Supplementary F for more details.   

\begin{rem}
The purpose of this example is to convey the message that our general theory of covariate-adjusted rEB \emph{reduces} to conventional regression-adjusted empirical Bayes model (``first-order'' EB model) when the relevance function $d_x(y) \equiv 1$, and customizes non-parametrically through $d_x(y) $, otherwise. 
\end{rem}

\begin{figure} 
    \centering
    \vskip2em
    \includegraphics[width=.92\linewidth,trim=1.5cm 0cm 2cm 1.5cm]{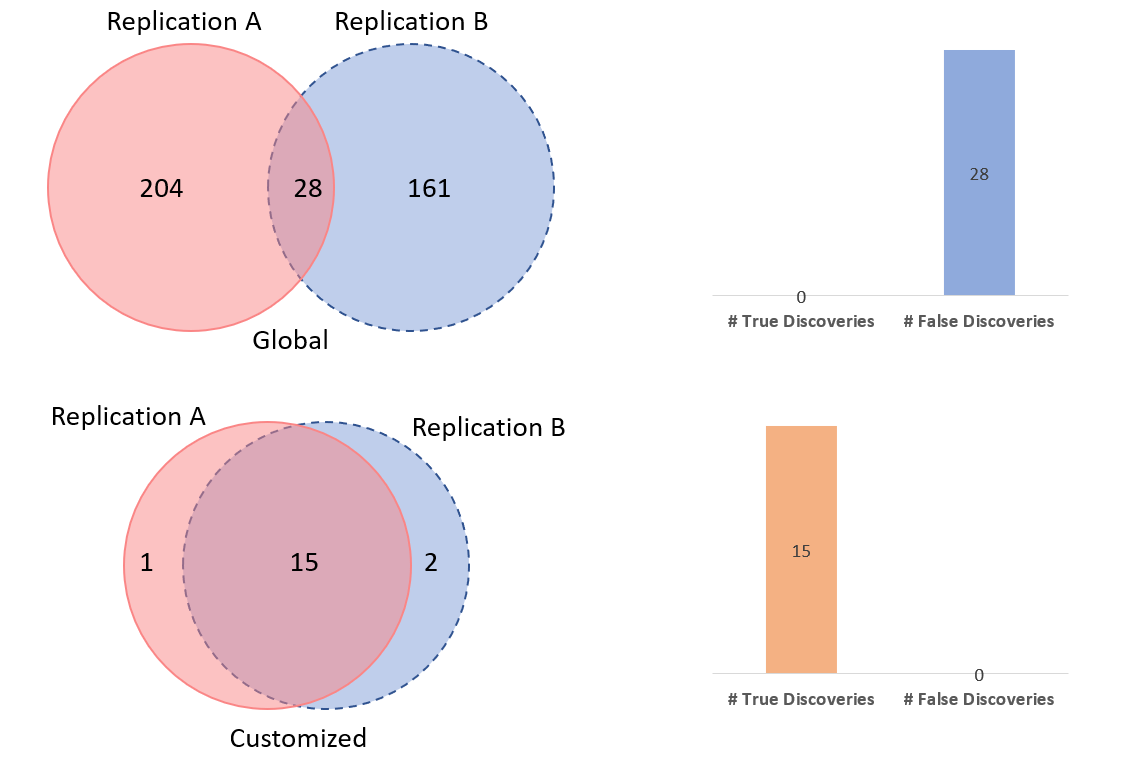}
    \vskip1.2em
    \caption{(color online) Two independent replications were done (they were generated independently from the same model using two different random seeds). Top panel: global methods have only $28$ discoveries in common, and shockingly all of them are false! Lesson learned: reproducible discovery $\not\Rightarrow$ correct inference. Bottom panel: the relevance-integrated customized methods are more reproducible; all the common $15$ discoveries are true signals.}
    \label{fig:REPRO}
    \vspace{-.5em}
\end{figure}
    
\vspace{-.2em}
\subsection{Robust Reproducible Inference} \label{sec:repro}
It is shown here that ignoring relevance, not only reduces the performance of an inference algorithm, but also drastically exacerbates the replicability crisis. Consider Fig. \ref{fig:REPRO}, which is based on two independent replications (i.e, we have used two different random seeds to generate the datasets from the same model or phenomena) of the \texttt{funnel} problem. For each one of the datasets, we applied both global and customized large-scale testing procedures. The top panel shows the results of the global locfdr method, where we have very few discoveries in common. But the most alarming part is that among these $28$ reproduced discoveries, not a single one is a true signal! This is shocking, to say the least. By contrast, the LASER-guided locfdr method shows exceptional performance: it finds all the $15$ effects that are ``reproducibly significant.'' 

\begin{rem}
The moral of the story is this:
\vskip.2em
~~~1. \textit{Relevance is underrated}: When replicated experiments do not yield the same results, should we panic and call it a `crisis' or figure out why? Modern experiments are complex and delicate, with several unknown moving parts. To ensure reproducibility, we must apply the principle of relevance to design \textit{robust large-scale inference} methods that can efficiently withstand unknown heterogeneity shocks. Robustness is the key to reproducibility.
\vskip.5em

~~~2. \textit{Reproducibility is overrated}: Reproducibility by its own does not serve as a ``stamp of approval'' to a correct inference, especially in the presence of heterogeneity ---commonplace in genomics and neuroscience. The dual objective of ``Relevance + Reproducibility'' seems to be a better goal to strive for.
\end{rem}
\subsection{A Universal Converter}
Here we discuss some practical benefits of our proposed customized-inference framework, which proceeds as follows: (i) Choose an appropriate global inference model (\textit{any} large-scale testing or estimation method); (ii) generate \texttt{LASER}($N;x$) by estimating the relevance function $d_x(z)$; (iii) feed those LASERs into the selected global model to individualize the inference based on case-specific characteristics.

This modular architecture makes the computational interface extremely simple and robust. If, in the future, we want to change, upgrade, or add new global inferential procedures, none of these changes will affect the LASER-based individualization process --- since we do not have to redesign the customization principle every time separately for each algorithm. This makes the whole implementation pipeline easily adaptable and specializable, which could be helpful for applied researchers and data scientists.


\section{Discussions}
\begin{quote}
``\textit{We are only beginning to recognize the many roles of borrowing strength.
We need to do this more rapidly, more widely and in more diverse situations.}'' (Mallows and Tukey, 1982)  \nocite{mallows1982}
\vspace{-.4em}
\end{quote} 
Statistical inference is a problem of `learning by comparison.' To tackle real-life modern statistical inference problems, we have to face the question: \textit{how to compare a large number of heterogeneous parameters in a meaningful way}? The key obstacle to addressing this question lies in the difficulty of resolving the ``relevance paradox,''  without proper consideration of which, even a prudent statistical inference method can go awry. This paper offers the first practical theory of relevance (with precisely describable statistical formulation and algorithm) to extract individual-level customized inference from increasingly massive and heterogeneous data sets. The advocated robust large-scale inference technology offers a simple mantra: ``personalize your inference by feeding LASERs into your global full-data-based models.'' 
It is our hope that this simple and general principle will take us close to the ultimate goal of building an inference machine with contextual adaptation, which could be a powerful tool for precision medicine, healthcare,  recommendation system, defense, and national security-related applications.

\section*{Acknowledgement}
This paper is dedicated to the ``50 Years of the Relevance Problem'' --- a long-neglected topic that begs attention from practical statisticians who are concerned with the problem of drawing inference from large-scale heterogeneous data.

The authors would like to thank the editor, associate editor, and  the two anonymous reviewers for their clear and concise suggestions. This research was inspired by a question raised by Brad Efron to one of the co-authors. We are grateful to  Brad  Efron and  Jerry  Friedman for thoughtful discussions and valuable comments.

\section*{Supplementary material}
\label{SM}
The online supplementary material includes additional methodological and numerical details. The proposed relevance-integrated inference procedure (including all the datasets) is implement in the \texttt{R} package \texttt{LPRelevance} \citep{LPRelRpck}. Available online: \mbox{https://CRAN.R-project.org/package=LPRelevance}. \nocite{LPRelRpck}

\bibliographystyle{Chicago}
\bibliography{ref-bib2}

\newpage
\clearpage
\setcounter{figure}{0}
        \renewcommand{\thefigure}{S\arabic{figure}}
\setcounter{table}{0}
        \renewcommand{\thetable}{S\arabic{table}}
\begin{center}
{\Large {\bf Supplementary Material for\\[.4em] `On The Problem of Relevance in Statistical Inference'}}\\[.5in] %
\end{center}

\begin{figure}
        \centering
     \includegraphics[height=.175\textheight,trim=.5cm .5cm .5cm .5cm]{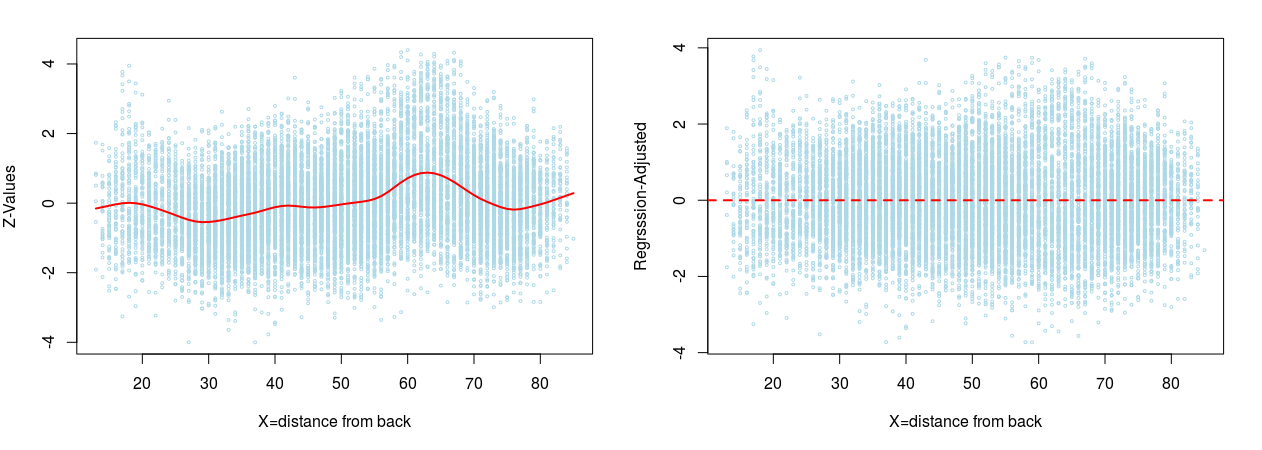}\\[.45em]
      \includegraphics[height=.175\textheight,trim=.5cm .5cm .5cm .5cm]{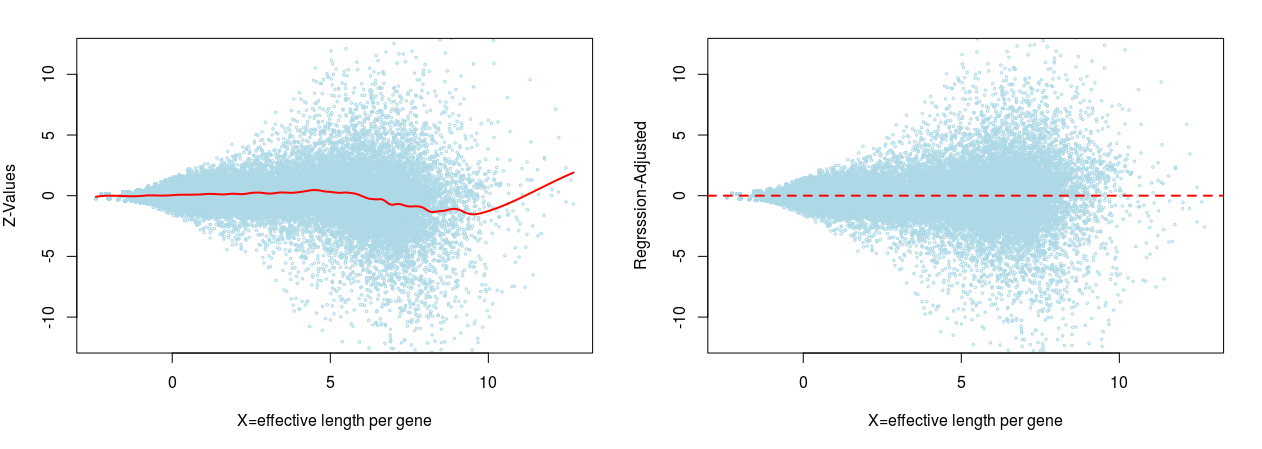}\\[.45em]
       \includegraphics[height=.175\textheight,trim=.5cm .5cm .5cm .5cm]{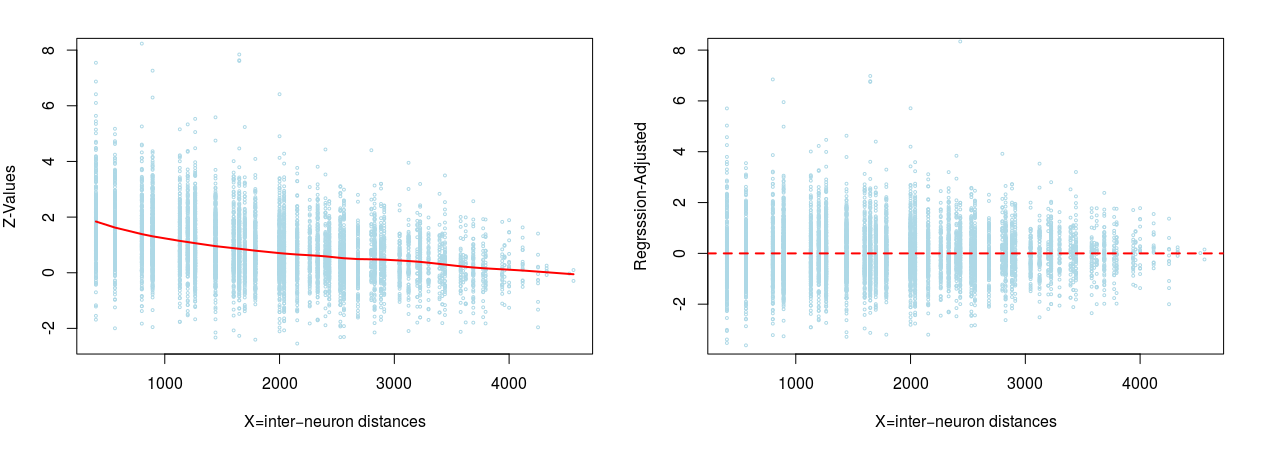}\\[.45em]
        \includegraphics[height=.175\textheight,trim=.5cm .5cm .5cm .5cm]{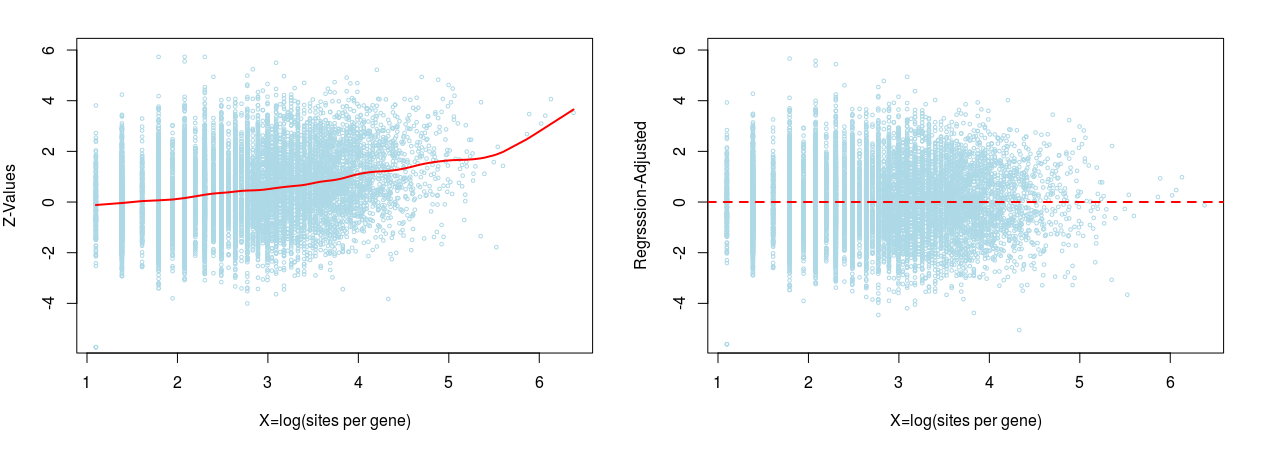}
        \vskip.55em
    \caption{Four real datasets are shown in different rows. First row: DTI data as described in the section \ref{sec:app1} of our paper; second row: airway smooth muscle cell lines RNA-Seq data \citep{himes2014rna}; third row: a neuroscience experiment to detect interactions among 128 neurons recorded simultaneously from the primary visual cortex (V1) of a rhesus macaque monkey \citep{kelly2007}; forth row:  Chi-square genomics data \citep[p. 92]{efronlsia}. Left columns show the $(x,z)$ plots and the right columns shows the regression-adjustment plots. It is clear that the naive first-order heterogeneity correction is not sufficient. Unfortunately, most of the current large-scale inference methods deal heterogeneity just by adjusting the conditional mean and thus, often leads to erroneous conclusions like Fig. \ref{supp:comp}.}
\label{supp:4real}
     \end{figure}

This supplementary document contains eight Appendices. 
\begin{center}
{\large A. Large Shallow Heterogeneous Data: Real Examples}
\end{center}
To understand how the \texttt{funnel} data-like situation can routinely arise in practice, here we discuss four real-datasets. The left panel of Fig. \ref{supp:4real} shows the original $(x,z)$ plots and the right panel shows the regression-adjusted ``flattened'' plots. All of these examples share two common characteristics (apart from being large):
\begin{itemize}[topsep=2pt]
  \setlength{\itemsep}{1.65pt}
    \item Shallow:  They are of shallow depth--at each covariate value $x_i$, we have very few $z$'s.  
    \item  Excess Heterogeneity: Waves of heterogeneity exists beyond (first-order) regression-adjustment. Many large-scale biomedical studies (e.g., https://allofus.nih.gov/) purposely incorporate participants with different races, ethnicities, age groups etc. For investigators, understanding ``diversity'' (heterogeneity) is of prime importance to enable individualized treatment.  
\end{itemize} 

To move from `one-size-fits-all' global inference to case-specific `precision inference,' we must address the real question: how to effectively deal with \textit{distributional} heterogeneity (beyond the obvious mean-effect)? The  \texttt{funnel} data is designed to facilitate this discussion.

Conventional large-scale inference methods primarily inspect: \textit{which} cases are significantly large (by developing sophisticated p-value thresholding rule)? But, that could be a highly misleading question when heterogeneity is present as they are not directly comparable. We have to also ask \textit{what} makes them large: is it because of the large variability (or even long-tailedness) at certain $x=x_0$ which gave them a `lucky' push or they truly outrank other `similar' cases. By ``similar' we mean cases with comparable heterogeneity (statistical fluctuations) level. To separate `real' signals from the `cosmetic' ones, we have to tame the heterogeneity, and running a regression-smoother is just the first step, not the last. 

\begin{center}
{\large B. Comparison With Other Global/Semi-Global Methods}
\end{center}
\vspace{-.4em}
The task of separating signals from unknown background variation is a delicate business (discovery $\neq$ application of multiple-testing procedures on the p-values; nothing is `absolute,' not even p-values, it depends on the choice of a ``relevant null''; see Appendix D), whose success critically depends on ``appropriately'' choosing relevant comparison samples for each case individually. 
     
\vskip.8em
\begin{figure}[ht]
        \centering
     \includegraphics[width=\textwidth,trim=1cm .5cm 1cm 1cm]{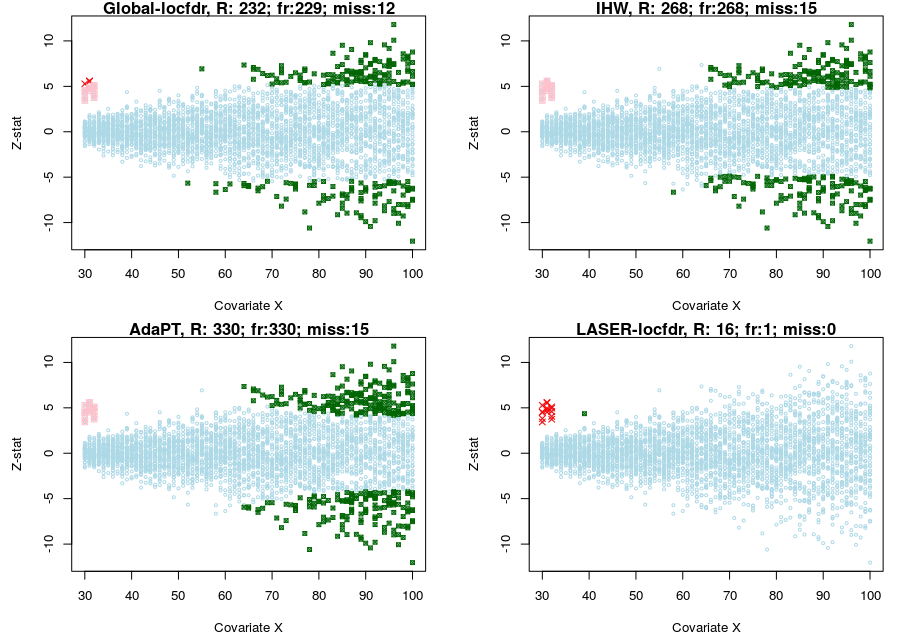}
\vskip1em
    \caption{Comparison of four methods for the \texttt{funnel} data. Top left is the global locfdr; top right and bottom left are respectively IHW and AdaPT; bottom right is our customized method that is `locfdr+LASER'. `R' stands for number of rejections, `fr' means number of falsely declared signals, and `miss' denotes number of true signals missed.} \label{supp:comp}
     \end{figure}
Fig. \ref{supp:comp} compares 4 methods: global locfdr, IHW \citep{Ig2016}, AdaPT \citep{AdaPT18}, and our customized locfdr. In the light of Theorem 3, global completely ignores the ``relevance'' correction” part inside the square brackets of \eqref{eq:fdrexp}. Whereas the semi-global methods like AdaPT and IHW assume that the null distribution is not changing with the covariate ($X \bot Z$ under $H_0$): $f_0(z|x)=f_0(z)$, which completely sabotages their usefulness. On the other hand, LASER-guided locfdr shows impeccable performance. The important point to note here is that: we have not designed any new specialized version of locfdr, instead we have just changed the diet of the global locfdr by feeding LASERs into it. 

\begin{center}
{\large C. Effective Relevant Sample Size [rESS]}
\end{center}
\begin{quote}
\vspace{-.3em}
``\textit{Relying entirely on direct evidence is an unaffordable luxury in large-scale data analyses, but indirect evidence can be a dangerous sword to wield. Some theoretical guidance would be welcome here, perhaps a theory quantifying the relevance of group data to individual estimates.}''~(Efron, 2010)
\end{quote}
\vspace{-.3em}
\begin{figure}[!ht]
\vskip.2em
    \centering
    \includegraphics[width=.65\linewidth,trim=1cm .5cm 1cm .5cm]{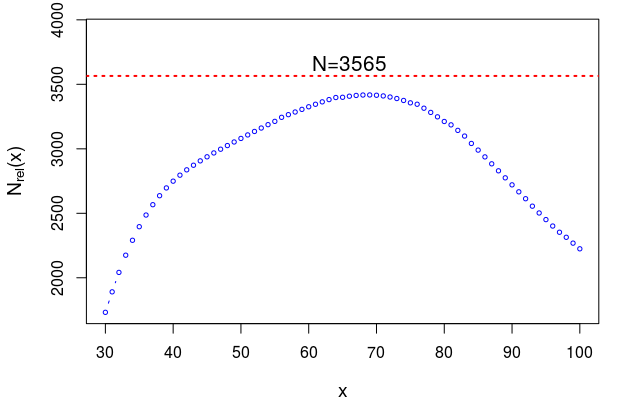}
    \vspace{-.2em}
    \caption{The effective relevant sample sizes (\texttt{rESS}) for \texttt{funnel} data. In terms of efficiency, the cases with small values of $x$ suffer the most. On the other hand, the cases at the center (around $x=65$ or so) enjoy the luxury of utilizing nearly all of the direct samples as relevant set. This prompts the question: can we synthesize (instead of searching) $N=3,565$ relevant learning samples for each case? This will take care the inherent efficiency/relevance bias (due to unequal \texttt{rESS}) originating from the underlying heterogeneity. This is the motivation behind LASERs, which is discussed in Section \ref{sec:sampler} of the main paper.}
    \label{fig:Nrel}
\end{figure}
The diagnostic plot in Fig. \ref{fig:Nrel} implements the formula \eqref{eq:rel} of the main paper. It shows the effective number of relevant cases as a function of $x$.

\begin{center}
{\large D. Relevant Null Estimation: Two Approaches}
\end{center}
\vspace{-.4em}
For a target group with $X=x$, we are interested in estimating the parameters of the relevant empirical null $\cN\big(\mu_0(x), \sigma_0(x)\big)$. Not that this is not directly computable since we have too few cases at a particular $X=x$. Here we discuss two approaches as to how we can tackle this problem.

\vskip.6em
{\bf Approach 1}. The first approach is exceptionally simple, consisting of two steps: First, simulate \texttt{LASER}($N;x_0$) from the big heterogeneous data by applying Algorithm 1 (Section \ref{sec:sampler}); Then, use any global empirical null estimation procedure on the \texttt{LASER}($N;x$) to get the parameters $\mu_0(x)$ and $\sigma_0(x)$. Our R-package \texttt{LPRelevance} currently implements locfdr \citep{efron04a} and QQnull \citep[Sec. 3.5]{deep16LSSDBio} as the two choices for estimating global empirical null. This process is illustrated in Fig. \ref{fig:rebnull}.

\begin{figure}[!ht] 
\vskip.2em
    \centering
    \includegraphics[width=.45\linewidth,trim=1cm 1cm 1cm 1cm]{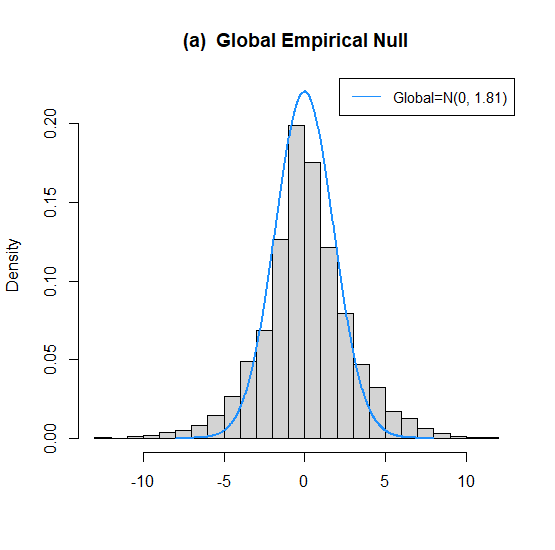}~~~~~
     \includegraphics[width=.45\linewidth,trim=1cm 1cm 1cm 1cm]{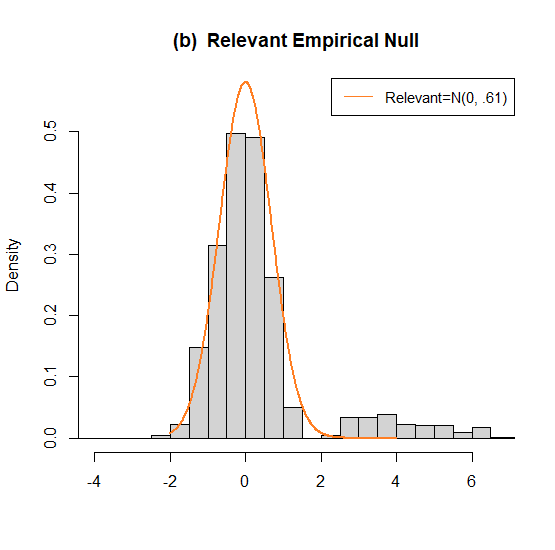}\\[1em]
     \includegraphics[width=.55\linewidth,trim=1cm 1cm 1cm 1cm]{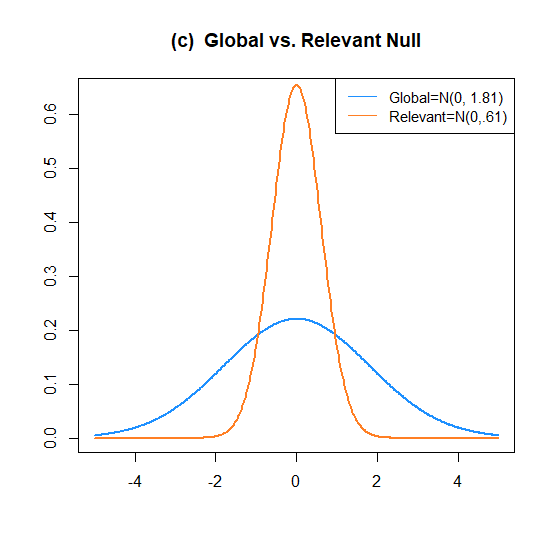} 
     \caption{We compute and compare the shapes global and relevant empirical (at $X=30$) nulls for the funnel data. (a) The global empirical null computed by applying locfdr() routine on the full data; (b) The histogram is the laser samples (at $x=30)$ on which we applied the locfdr() routine to get the relevant null; (c) It contrasts two nulls to underscore the danger of assuming $f_0(z|x)=f_0(z)$ in practice.}
\end{figure}
\begin{figure}
\vspace{-.8em}
    \centering
    \includegraphics[width=.6\linewidth,trim=1.2cm .5cm 2cm 1cm]{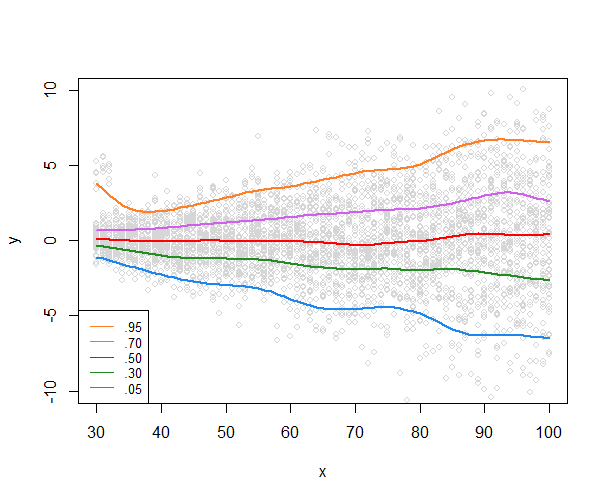}
    \caption{The estimated conditional quantile curves using our global-to-local conditional density model.}
    \label{fig:funnelQR}
    \vskip.4em
\end{figure}

 \label{fig:rebnull}
\vskip.6em
{\bf Approach 2}. An alternative way to estimate the conditional null parameters is via quantiles. First, we estimate the conditional quantile functions $Q(u|x)$. This can be done easily using our global-to-local model of Sec. \ref{sec:g2lm} as shown in Fig. \ref{fig:funnelQR} for the \texttt{funnel} data. Next, we estimate the relevant null parameters simply by:
\beas \widehat \mu_{0}(x)&=&\whQ(.5|x)\\[.5em]
\widehat \si_{0}(x)&=& \frac{\whQ(.75|x)- \whQ(.25|x)}{1.349}
\eeas
Now note that, from Eq. \eqref{eq:g2l} we immediately have:
\beas 
F(z|x)&=& \int_{-\infty}^{z} f(t) \,d(F(t);Z,Z|X=x) \dd t\\[.5em]
&=&\int _0^{F(z)} d(u;Z,Z|X=x) \dd u~~~\text{(by substituting $F(t)=u)$}\\[.5em]
&=&D\big(F(z); Z,Z|X=x\big),
\eeas
where $D(u;Z,Z|X=x)$ is the cdf of the relevance density function $d(u;Z,Z|X=x)$. As a consequence, we can now express the whole correction factor \eqref{eq:fdrexp} solely as a function of relevance function $d_x$.

{\bf Note}. Classical theory of simultaneous inference heavily relies on p-values, i.e., on null hypothesis tail areas. Needless to say, the real value of p-value depends on what we pick as our null distribution $F_0$. (i) The usual practice is to take $F_0=\cN(0,1)$. We call it \textit{absolute} p-values. (ii) \textit{Empirical} p-values computed based on global empirical null \citep{efron04a} $\wtF_0=\cN(\tilde \mu_0,\tilde \sigma_0)$, as in Fig. \ref{fig:rebnull}(a). (iii) In contrast, the \textit{relevant} p-values (used in Fig 9 `LASER+BH') are computed using context-adaptive null $\wtF_{0|x}=\cN(\tilde\mu_{0|x},\tilde\sigma_{0|x})$, see Fig. \ref{fig:rebnull}(b).

\begin{center}
\vskip.5em
{\large E. Regarding DTI Data}
\end{center}

\begin{figure}
    \centering
    \includegraphics[width=.64\linewidth]{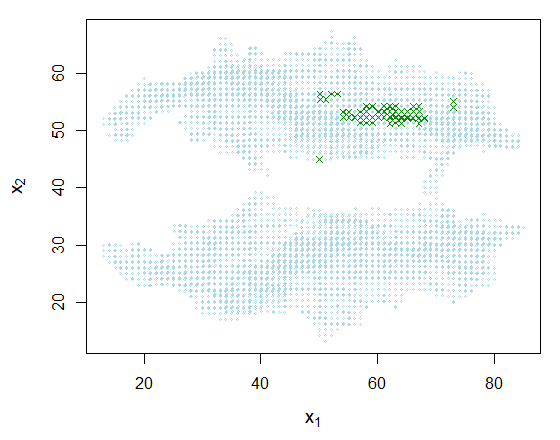}\\[1.4em]
    \includegraphics[width=.48\linewidth]{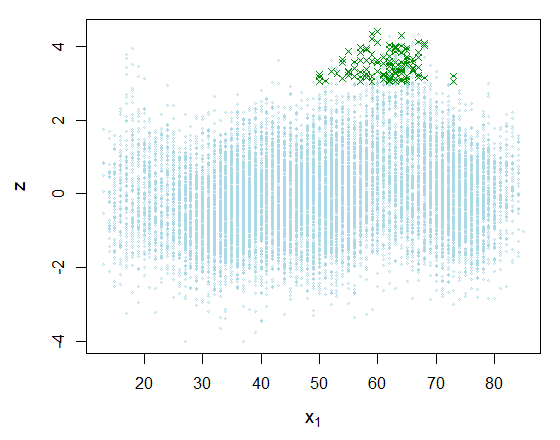}~~~
    \includegraphics[width=.48\linewidth]{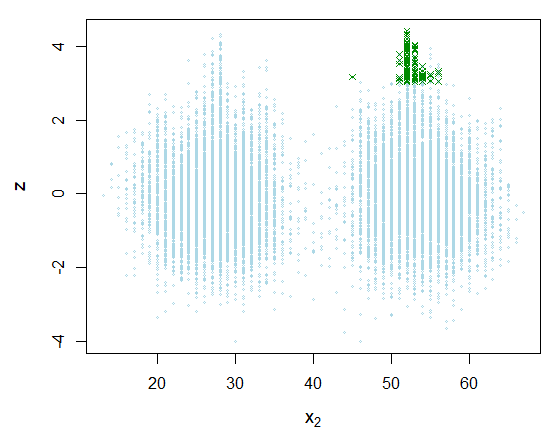}
    \vskip1em
    \caption{This plot complements Fig. \ref{fig:dti_macro} of the main paper. In green, we show a cluster of $111$ voxels---around the left frontal lobe, which were primarily declared as significant by the global locfdr method but have been tactically avoided by our relevance-integrated customized locfdr method. Here $x_1$ denotes the distance from the back, and $x_2$ distance from the right. Note that all of the green voxels concentrated near $x_1=60$ and $x_2=55$, which naturally creates suspicion that they are loud noise, not true signals--who got ``lucky push'' from heterogeneity. Our LASER-guided individually tailored-inferential scheme helps to separate these type of man-made signals from nature's hint.}  \label{fig:suppDTI}
\end{figure}

Fig. \ref{fig:suppDTI} provides further details on the findings of DTI data; see Section \ref{sec:app1} of the paper.

\newpage

\begin{center}
{\large F. Relevance-Integrated Empirical Bayes: A Step by Step Guide}
\end{center}

We illustrate the main steps of global empirical Bayes (gEb) and our relevance-integrated empirical Bayes (rEB) method in the context of the \texttt{funnel} data, Fig. \ref{fig:EB}.

\vskip.65em
{\bf F.1~ Global Empirical Bayes}. Standard EB practice proceeds in three steps: First,  compute regression-adjusted $y$ from $z$-values
\[y_i = z_i - \widehat{\Ex}[z \mid x_i],~~i=1,2,\ldots,n.\]
Secondly, learn the global EB prior $\widetilde{\pi}_G(\te)$ (same for \textit{all} cases) from the full data $y_1,\ldots,y_N$; and finally,
compute the posterior $\widetilde{\pi}_G(\te|z_A)$ for case A based on the global prior. This is displayed as blue dotted line in the bottom two panels of Fig. \ref{fig:EB}.

\vskip.65em
{\bf Note 1}. We have used the method of `DS-Prior,' proposed in \cite{deep18nature_app} to \textit{learn} the global prior from large number of parallel cases. We prefer this approach because of its stability, flexibility, and robustness. The other advantage of the DS-prior is that it produces a closed-form parameterizable $\widehat \pi$ that facilities easy simulation.   The method is implemented in the R-package \texttt{BayesGOF} \citep{BayesgofR_app}.

\vskip.65em
{\bf F.2~ Relevance-Integrated Empirical Bayes}. To perform rEB analysis for case A with $x=x_A$ and $z=z_A$, we proceed as follows:

\vskip.3em
\texttt{Step 1.} Estimate the relevance function $\whd_{x_A}(z)$ and plug this into Algorithm 1 to generate the \texttt{LASER}($N,x_A$).
\vskip.3em
\texttt{Step 2.} Apply your favorite global EB prior estimation algorithm on the lasers to get the custom-prior 
$\widehat{\pi}_A(\te|x_A)$ for case A. It is instructive to compare the shapes of global (blue) and relevant priors (red) in Fig. \ref{fig:EB}. The global EB-prior is much more diffused, whereas the rEB prior is considerably more peaked around zero and has a bump around $\te=2.50$.

\vskip.3em
\texttt{Step 3.} The rEB posterior distribution $\widehat{\pi}_A(\te|x_A, z_A)$ is shown in the right plot of middle panel. Note the striking differences between the global (blue) and the rEB posterior (red).
\vskip.4em
{\bf Note 2}. The context-adaptive rEB prior was instrumental in accurately estimating the effect size parameters by controlling two types of biases: selection bias and relevance bias. It is thus important to give meticulous attention in constructing relevant-EB prior that faithfully captures and represents the background context of the target case A. 
\vskip.4em

\texttt{Step 4.} [Optional] To boost the quality of the inference further (at the cost of more computation) repeat the process $B$ times ($B=10$ is often good enough) and return the averaged-prior and posterior. All of these can be done easily using the \texttt{rEB.proc}() function.

\vskip.25em
\begin{lstlisting}
#rEB microinference for case A
> library(LPRelevance)
> rEB_A<-rEB.proc(X,z,X.target=30,z.target=4.49,niter=10)
> rEB_A$result #prior and posterior summaries
> rEB_A$plots$rEB.prior  # Fig 13: left plot in the middle panel
> rEB_A$plots$rEB.post # Fig 13: right plot in the middle panel
\end{lstlisting}
\vskip1em


{\bf F.3~ Empirical Bayes Confidence Interval Estimate}. A robust nonparametric method for EB confidence interval construction is an important (not yet fully solved) problem. We tackle it as follows:

\vskip.4em
\hskip7em Algorithm 3. ~~Construction of Finite Bayes rEB CI
\vspace{-1.24em}

\rule{\textwidth}{.8pt}
\vskip.1em

\texttt{Step 0.} Using sampling Algorithm 1, 
simulate \texttt{LASER}($N,x_A$). Compute the standard deviation
$\hat \sigma_A$ by taking the  IQR/1.349 of those laser samples.

\vskip.55em
\texttt{Step 1.} Estimate the rEB DS-prior $\widehat{\pi}_A(\te|x_A)$ from the laser samples. 

\vskip.55em
\texttt{Step 2.} Simulate $\te=(\te_1,\ldots,\te_N)$ from the estimated prior $\widehat{\pi}_A$ following the recipe given in \citet[Supp. Appendix C]{deep18nature_app}.
\vskip.55em
\texttt{Step 3.} Generate the observation $y^*_i$ from $\cN(\te_i,\hat\sigma_A)$ for $i=1,2,\ldots,N$.
\vskip.55em
\texttt{Step 4.} Estimate the rEB prior $\hat \pi_*(\te)$ and the posterior  $\hat \pi_*(\te\mid y_A,x_A)$ based on ${\boldsymbol y}^*=(y^*_1,\ldots, y^*_N)$.
\vskip.55em
\texttt{Step 5.} Repeat steps 2-4 a large number of times, say $B=100$ times. Compute the averaged bootstrap rEB posterior distribution:
\beq \widecheck\pi_A(\te\mid y_A,x_A) \,=\,\frac{1}{B}\sum_{k=1}^B \hat \pi_*^{(k)}(\te\mid y_A,x_A).~~~\eeq
Return the $100(1-\al)\%$ rEB confidence interval (CI) for $\widehat\te_A$ by computing the credible interval
from the posterior distribution $\widecheck\pi_A$. The finite Bayes rEB posterior analysis can be done simply by calling the \texttt{rEB.Finite.Bayes}() function of the LPRelevance R-package. 
\\
\rule{\textwidth}{.8pt}
\vskip.6em

{\bf Note 3}. \cite{efron2019apx} called this bagging-style parametric-bootstrap averaging approach as ``Finite Bayes Inference'' that takes into account the additional uncertainty due to estimation of
EB prior (see step 1) $\widehat{\pi}_A(\te|x_A)$ from finite sample. The credible intervals based on $\widecheck\pi_A(\te|x_A,y_A)$ yield a properly calibrated (``wider'') CI, attaining the desired coverage $1-\al$. 
\newpage

\begin{center}
{\large G. Relevance-Integrated Simultaneous Testing: A Step by Step Guide}
\end{center}

\begin{figure}[ht]
    \centering
    \includegraphics[width=.8\linewidth]{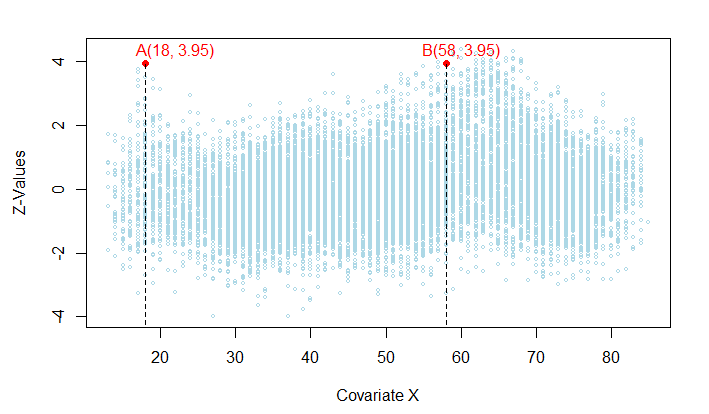}
    \caption{DTI data: $(z_i,x_i)$ for $i=1,\ldots,15,443$. We are interested in inferring about the cases A and B. Note that the voxels have the same $z$-value ($3.95$) but are situated at the different parts of the brain: one is at the back half and the other one is at the front half.}
    \label{fig:set_up}
\end{figure}

{\bf G.1 ~DTI data}. Following \cite{efron2019apx}, here we consider the following setup: For each of the $N=15,443$ voxels, we have two-sample $z_i$-statistic along with an additional piece of information\footnote[2]{The original bivariate case is discussed in Sec 3.3 of the main paper.} $x_i$, which says how far the voxel is from the back of the brain. Our goal is to perform microinference for case A and B, as shown in Fig. \ref{fig:set_up}.

\begin{figure}[t]
    \centering
    \includegraphics[width=.48\linewidth]{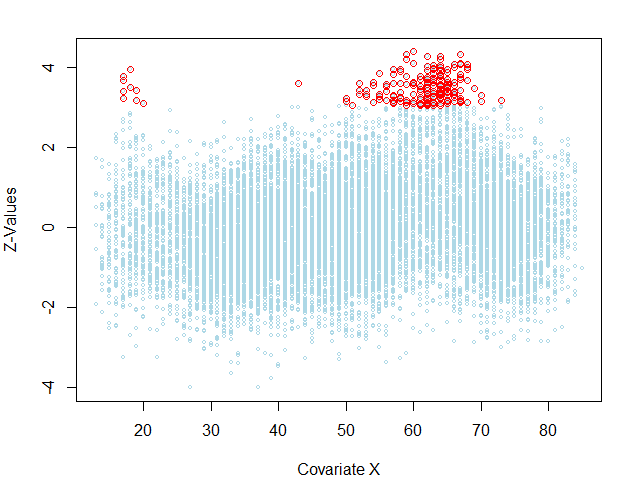}~~~
     \includegraphics[width=.491\linewidth]{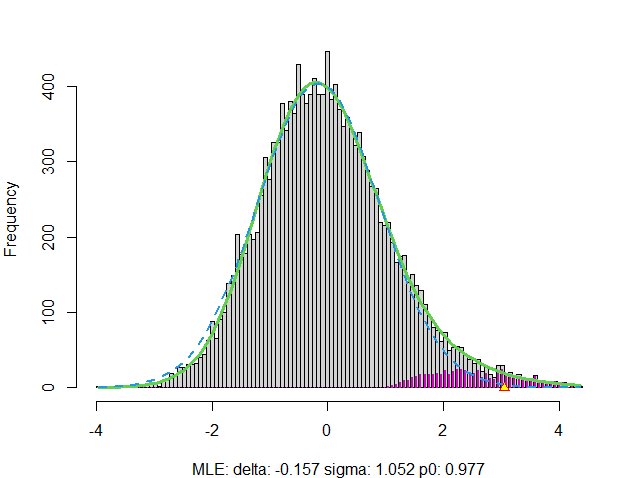}
     \vskip1em
    \caption{Global fdr analysis: Red dots are the $184$ significant voxels selected by locfdr.}
    \label{fig:global}
\end{figure}

{\bf G.2 ~Global Analysis}.  Fig. \ref{fig:global} reports the global fdr analysis, which finds both cases A and B to be significant (fdr $<0.2$):
\beq {\Pr}_{G}({\rm null} \mid z=3.95) = 0.034.\eeq

The caveat, of course, is that global analysis assumes we are given a large number of homogeneous (comparable) cases. In particular, it assumes that all the $z_i$'s are \textit{equally} relevant for assessing the significance of case A or B, which is clearly untrue. To allow customization, we have to think beyond the classical global inference scheme.

\begin{figure}[ht]
    \centering
    \begin{subfigure}{.465\linewidth}
        \caption{Back-half}
        \includegraphics[width=\linewidth,trim=1cm 1cm 0cm 0cm]{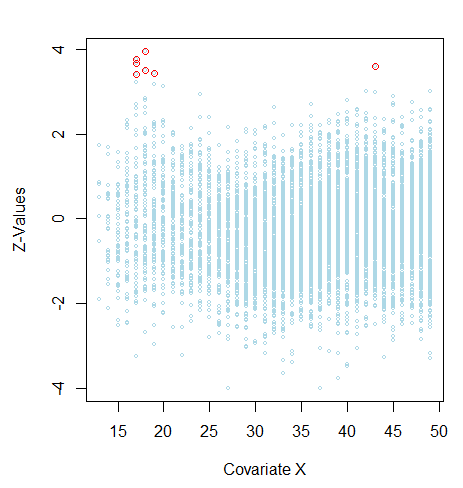}
    \end{subfigure}~~~
    \begin{subfigure}{.465\linewidth}
        \caption{Front-half}
        \includegraphics[width=\linewidth,trim=0cm 1cm 1cm 0cm]{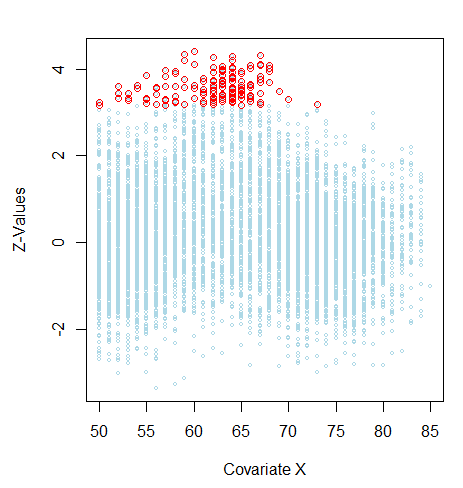}
    \end{subfigure}
    \vskip1em
    \caption{Separate analyses declared $7$ voxels from back-half (out of $N_0=7,661$) and $141$ voxels from the front-half (our of $N_1=7,782$) significant. MLE empirical null estimates for the  back-half and front-half voxels are respectively: $(-0.32, 0.98)$ and $(0.04, 1.09)$.}
    \label{fig:app1_dti}
\end{figure}

\vskip.6em
{\bf G.3 ~Semi-Global Subgroup Analysis}. 
\cite{efronlsia} took the first step towards tackling heterogeneity by partitioning the data into two groups: namely back ($x<50$) and  front halves ($x\geq 50$). The fdr was then computed \textit{separately} for the two brain regions. Fig. \ref{fig:app1_dti} shows the result, which finds $148$ voxels to be  significant (with fdr $<0.2$): $7$ came from the back-half and the remaining $141$ came from the front-half; also see Fig. 10.1 of \cite{efronbook}. Both cases A and B were picked as significant voxels with 
\bea 
&\fdr_A(z|I=0)\,=\, \Pr_A({\rm null}|z=3.95,I=0) \,=\, 0.07, \label{eq:p1} ~~~\\
&\fdr_B(z|I=1)\,=\, \Pr_B({\rm null}|z=3.95,I=1)\, =\, 0.06. ~~~\label{eq:p2} 
\eea
where $I$ is a indicator variable denotes back and front-half of the brain region, .i.e., $I=\mathbb{I}\{x\geq 50\}$.

\vskip.8em
{\bf G.4 ~  Heterogeneity and Relevance Diagnostic}.  Can we borrow strength from all of the $N_0=7,661$ voxels located in the back of the brain  to make precise inference about case A? If all of the $N_0$ voxels are equally relevant for case A, then  the answer is yes. Otherwise (i.e., when the voxels are heterogeneous), naively borrowing information might deteriorate the quality of the inference. Define the set $\mathbb{Z}_l$ which consists of z-values with $I_i=l$ for $i=1,2,\ldots,N$ and $l=0,1$, with empirical cdf $\wtF_l$.

\begin{figure}[ht]
    \centering
     \begin{subfigure}{.47\linewidth}
        \caption{~$d^0_A(z)$}
        \includegraphics[width=\linewidth,trim=1cm 1cm 0cm 0cm]{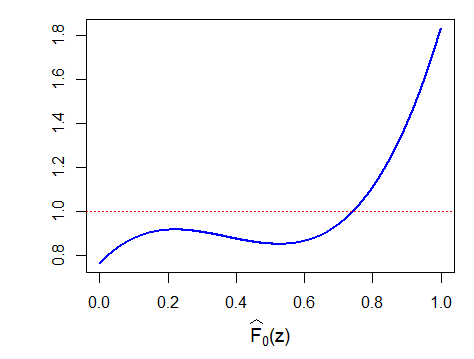}
    \end{subfigure}
    \begin{subfigure}{.47\linewidth}
        \caption{~$d^1_B(z)$}
        \includegraphics[width=\linewidth,trim=0cm 1cm 1cm 0cm]{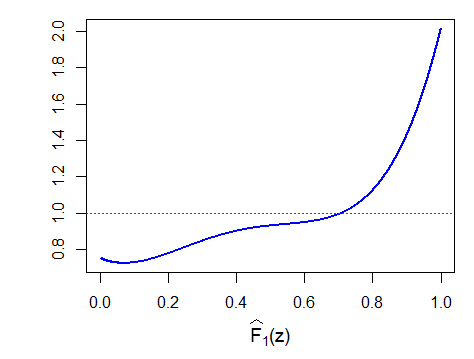}
    \end{subfigure}
    \vskip1.5em
    \caption{(a) relevance function for case A, with respect to data from back part of the brain; (b) relevance function for case B, with respect to data from front part of the brain. They act as a nonparametric exploratory diagnostic tool to check the appropriateness of the `separate' fdr analysis done in Sec G.3.}
    \label{fig:app_heterocheck}
\end{figure}

The `relevance question' then boils down to understanding how representative is $\mathbb{Z}_0$ for case A? A formal answer to this question is given by estimating the following relevance function
\beq d^0_A(z): =d(\wtF_0(z);\mathbb{Z}_0,\mathbb{Z}_0|X=18)=1+0.17 \wtT_1(z)+ 0.13\wtT_2(z)+ 0.08\wtT_3(z).\eeq
Similarly, for case B, we have:
\beq d^1_B(z): =d(\wtF_1(z);\mathbb{Z}_1,\mathbb{Z}_1|X=58)=1+0.24\wtT_1(z)+ 0.11\wtT_2(z)+ 0.08\wtT_3(z).\eeq
Shapes of these two relevance functions are shown in Fig. \ref{fig:app_heterocheck}. The non-uniformity of $d^0_A(u)$ indicates the $N_0$ back-half voxels are not equally relevant for case A. This calls into question the credibility of the semi-global analysis done in Sec G.3.  The same goes for the case $B$.

\vskip.8em
{\bf G.5 ~ What's the way forward?}  A naive attempt: Let's keep refining the partitions to get more and more homogeneous (comparable and relevant) cases to learn from. To do that efficiently, we have to decide on few things: how many partitions do we need?  how to select those partitions (where to cut)? These questions become especially challenging when we have multivariate $X$.  

The bigger issue with this line of investigation is that it will eventually lead us to the relevance paradox (see Fig. \ref{fig:relpara}), since as the number of partitions increases, the number of $z_i$'s for each partition will decrease, and hence the efficiency/reliability/computability of the large-scale inference procedures will drop drastically. All in all, this is a blind alley and we have to rethink our strategies.

Our global-to-local statistical inference model provides some practical (approximate) strategy on dealing with this impossible looking problem---which is described next.

\vskip.8em
{\bf G.6 ~ Proposed Method: LASER-guided Relevant Inference}. Our approach performs large-scale precision inference for voxels A and B. It has 4 main steps:

\vskip.65em
\texttt{Step 1.} \textit{Regression adjustment}. We start by taking out the obvious conditional mean heterogeneity by fitting user-specified regression function: $y_i = z_i -\widehat{\Ex}[z\mid x_i]$, for $i=1,\ldots,N$.

\vskip.65em
\texttt{Step 2.} \textit{Obtaining LP coefficients: Towards Relevance Function}. Question: whether after the regression adjustments the cases at $x=18$ and $x=58$ are comparable with the full data $y$-values.

To answer this question we estimate the relevance function $d_{x=18}(y)$ which compares the local conditional density $f(y|x=18)$ with the global marginal $f(y)$.
\beq d_x(y)= 1 + \sum_j\LP_{j|x} \wtT_j(y).\eeq
Following Theorem 1 (Section 2.3 of our main article), we estimate LP-coefficients via regressing:
\beq \widehat \LP_{j|x} \,= \,\widehat \Ex[  \wtT_j(Y) |X=x ]. \eeq
This is obtained by fitting a regression of $\wtT_j(Y)$ on $T_X$; see Section 2.3 and Theorem 2 for more details. Similar steps can be followed to get the coefficients of the relevance function $d_{x=58}(y)$. Fig. \ref{fig:app5_2_dtiTy} describes the regression method for obtaining the LP-coefficients,  which gives the following expressions for the relevance functions:
{\small
\beas 
d_{x=18}(y)&=&1-0.05\wtT_1(y)+0.07 \wtT_5(y)+0.09\wtT_6(y),\\  d_{x=58}(y)&=&1-0.03\wtT_1(y)+0.17 \wtT_2(y)+0.07\wtT_3(y)+0.11\wtT_4(y)+0.04 \wtT_5(y)+0.05\wtT_6(y).~~
\eeas
}

\texttt{Step 3.} \textit{Relevance Sampling:  Synthesizing LASERs}. Perform relevance sampling based on the $\whd_x(y)$, as detailed in Algorithm 1 of Section 2.5, and obtain the LASERs (in the $y$-domain) at $x=18$ and $x=58$, as detailed in the main paper and Supplementary D. One can easily switch to the $z$-domain by simply adding $\widehat{\Ex}[z|x_0]$ to the samples \texttt{LASER}($N;x_0$).

\begin{figure}[ht]
    \centering
    \begin{subfigure}{.455\linewidth}
        \caption{}
        \includegraphics[width=\linewidth,trim=1.5cm 0cm 0cm 0cm]{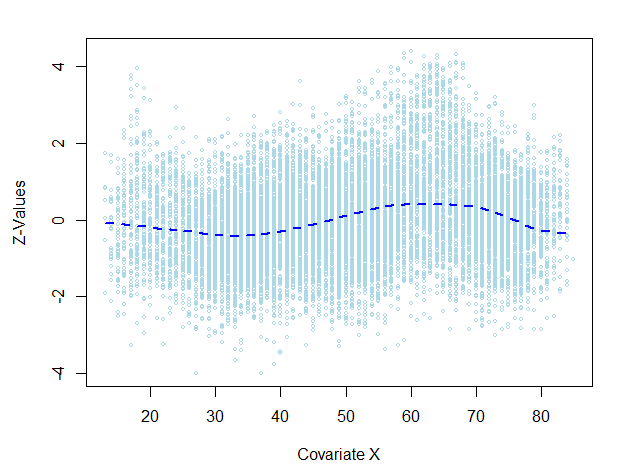}
    \end{subfigure}
    \begin{subfigure}{.455\linewidth}
        \caption{}
        \includegraphics[width=\linewidth,trim=0cm 0cm 1.5cm 00cm]{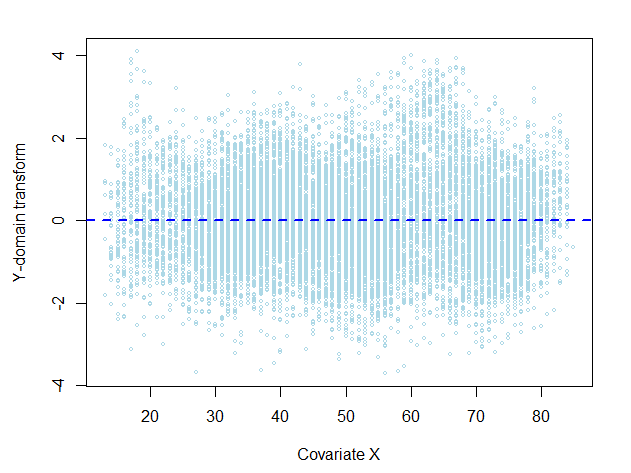}
    \end{subfigure}
    \caption{ (a) the DTI dataset with fitted regression curve, the fitted values are shown as the blue dotted line, which is to be subtracted to obtain $y$ values (b) Scatter plot of DTI data in $y$ domain, after removing the means in (a) from the original $z$ values.}
    \label{fig:app5_dti1}
\end{figure}

\begin{figure}[ht]
    \centering
    \includegraphics[width=.92\linewidth,trim=2cm 1cm 2cm 2cm]{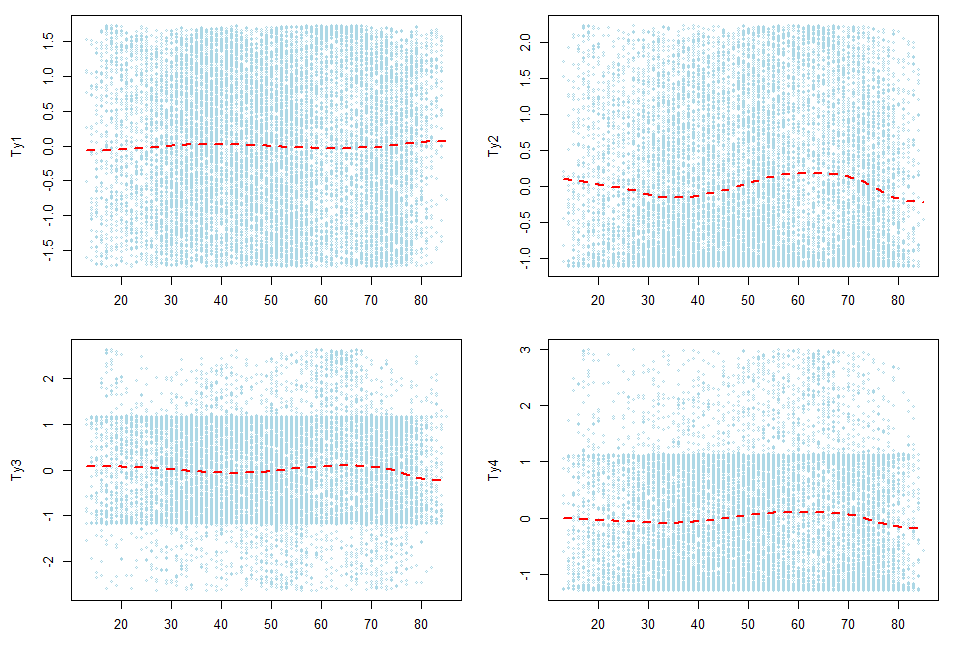}
    \vskip3em
    \caption{LP-regression($T_j(y) \sim X)$, for $j=1,\ldots,4$ reveal pattern beyond mean. The scatter plots show the transformed values of $T_j(y)$, while the dotted lines indicate the fitted regression line, whose values at $x$ are $\widehat{\LP}_{j|x}$ for $j=1,\ldots,4$.}
    \label{fig:app5_2_dtiTy}
\end{figure}

\begin{figure}[ht]
    \centering
    \includegraphics[width=.45\linewidth]{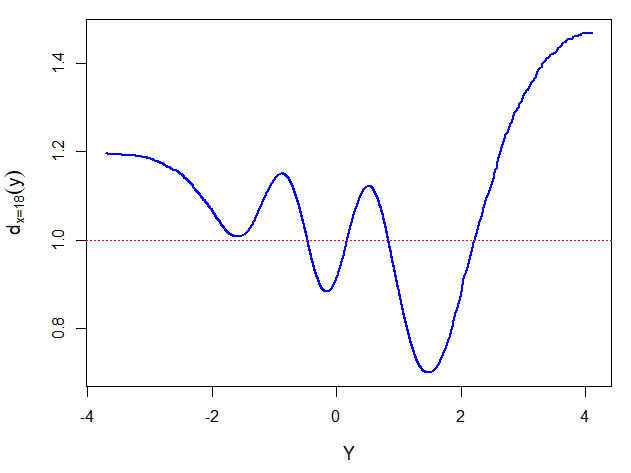}
    \includegraphics[width=.45\linewidth]{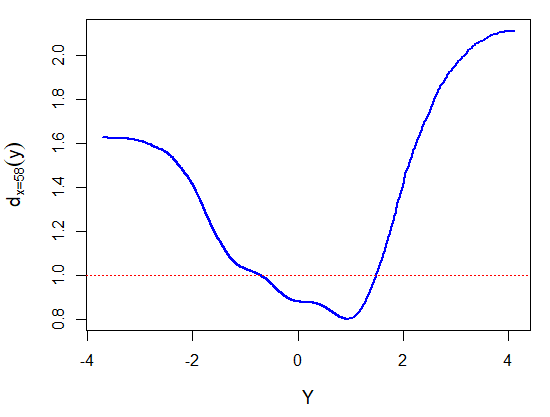}
    \caption{Relevance function for $x=18$ (left) and $x=58$ (right). Both indicating regression-adjustments is not enough to obtain homogeneous relevance samples.}
    \label{fig:app5_dti_lcoef}
\end{figure}

Fig. \ref{fig:app5_2_dtiTy} describes the regression method for obtaining the LP-coefficients, given below: 

\vskip1em
\begin{table}[ht]
    \centering
    \begin{tabularx}{.8\linewidth}{X X X X X X X}
    \toprule
         & $\LP_{1|x}$ & $\LP_{2|x}$ & $\LP_{3|x}$ & $\LP_{4|x}$& $\LP_{5|x}$ & $\LP_{6|x}$ \\
         \midrule
        Case $A$ & -0.05 &0 &0 & 0 & 0.07 &0.09 \\
        Case $B$ & -0.03 & 0.17 & 0.07 & 0.11&  0.04 & 0.05\\
        \bottomrule
    \end{tabularx}
    \caption{LP-coefficients (BIC-smoothed) at $x=18$ (case A) and $x=58$ (case B).}
    \label{tab:app5_coefx}
\end{table}

\vskip.8em
\texttt{Step 4.} \textit{LASER-guided Inference}.
\begin{itemize}[topsep=2pt]
    \item The relevant null construction:  $f_0(z|x)=\cN(\widetilde \mu_{0|x},\widetilde \sigma_{0|x})$, where $\widetilde \mu_{0|x}=\widehat \Ex[Z|X=x]$ and the null sd $\widetilde \sigma_{0|x}$ is estimated by applying  \texttt{locfdr} method on the $y$-domain LASER samples. The relevant empirical null distributions are shown in Fig. \ref{fig:app5_relnull}; contrasts the global empirical null with the relevant nulls at $x=18$ and $x=58$.
    \item The customized fdr curves are shown in Fig. \ref{fig:fdrplot}, from which we get:
    
    \bea
    \fdr_A(z|x=18)\,=\,{\Pr}_A(\textrm{null}|z=3.95,x=18)\,=\,0.04,\\
    \fdr_B(z|x=58)\,=\,{\Pr}_A(\textrm{null}|z=3.95,x=58)\,=\,0.19.
    \eea
   Contrast this with Eqs. \eqref{eq:p1} and \eqref{eq:p2} to note that our final conclusion is different from the subgroup-based analysis (see Sec G.3). In particular, we found strong evidence that voxel A is interesting, however, voxel B appears to be just a `loud noise' riding the wave of heterogeneity. 

\begin{figure}[ht]
    \centering
    \includegraphics[width=.6\linewidth,trim=2cm 1cm 2cm 2cm]{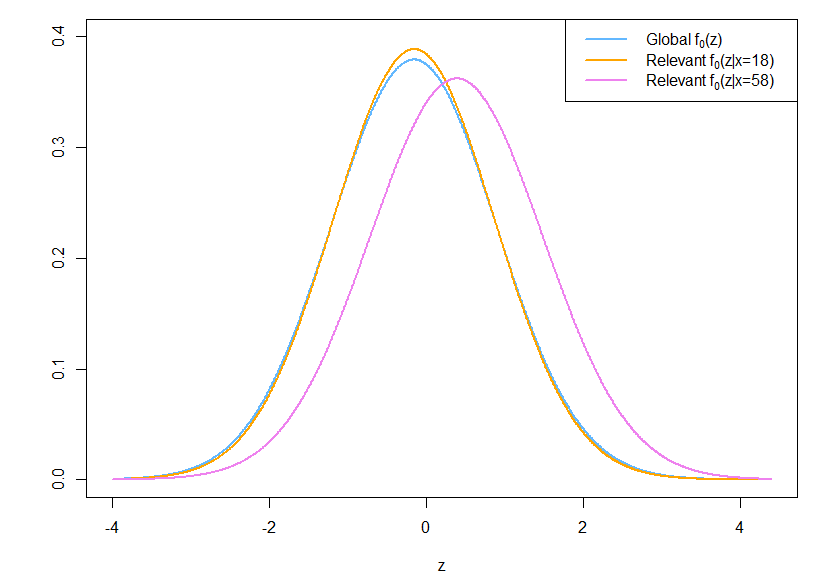}
    \vskip1em
    \caption{Comparison of global empirical null and relevant nulls computed via LASER samples.}
    \label{fig:app5_relnull}
\end{figure}

\begin{figure}[t]
    \centering
    \includegraphics[width=.475\linewidth]{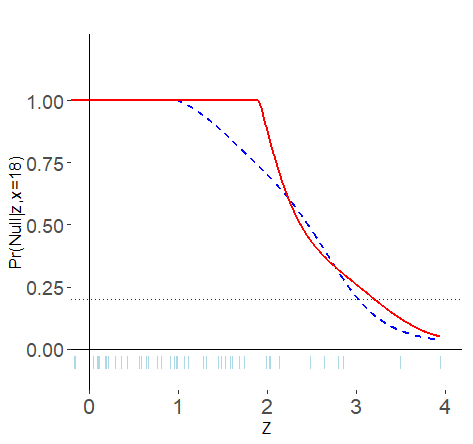}~~
    \includegraphics[width=.475\linewidth]{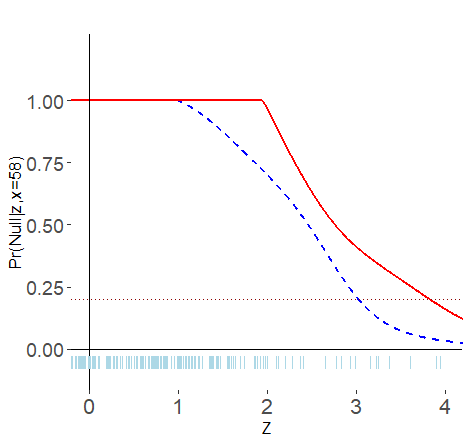}
    \includegraphics[width=.4\textwidth]{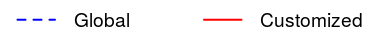}
    \caption{Customized fdr functions
    at $x=18$ (left) and $x=58$ (right), obtained by adjusting the global fdr curve (in blue).}
    \label{fig:fdrplot} 
\end{figure} 

\vskip1em
   {\bf Note}. Even though the two voxels have the same z value ($3.95$), our conclusions were different based on their backgrounds, as captured by the covariate $x$ (distance from the back). This shows the contextual adaptability property of our method, which is necessary to perform individualized inference from big heterogeneous data. 
    \item Following section 3.2, we have also performed the macro-inference, shown in Fig. \ref{fig:macroinf}.
\end{itemize}

\begin{figure}
    \centering
    \includegraphics[width=.7\linewidth,trim=1cm 0cm 1cm 0cm]{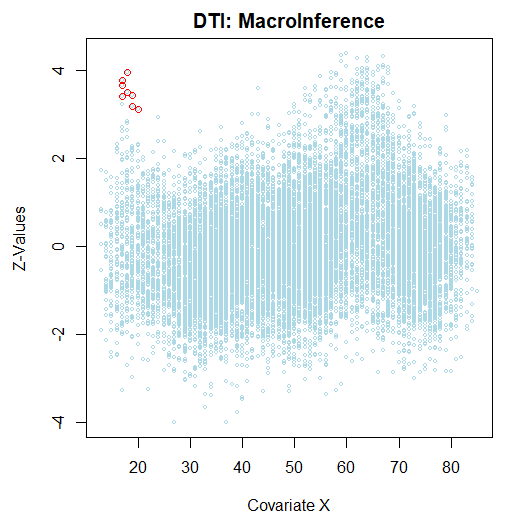}
    \caption{Macro inference results for DTI dataset using our proposed method: $8$ voxels (around $x=18$) are found to be interesting.}
    \label{fig:macroinf}
\end{figure}

\newpage

\begin{center}
\vskip.5em
{\large H.  Quantifying Uncertainty of Estimated $\whd_x(z)$ with Bootstrap}
\end{center}

Quantify uncertainty of the estimated relevance function $\whd_x$ by performing non-parametric bootstrap, which goes as follows:
\vskip.6em
\texttt{Step 1}. Resample with replacement from the original data: $(x^*_i,y^*_i)$, $i=1,2,\ldots,N$.

\vskip.4em

\texttt{Step 2}. Estimate the bootstrapped relevance function $\whd_x^*$ following the theory of Sec. \ref{sec:estrel}.
\vskip.4em

\texttt{Step 3}. Repeat steps 1 and 2, say $B=100$ times; compute the pointwise standard errors (or confidence bands) from the $B$ bootstrap replications.

\vskip1.55em
\begin{figure}[ht]
    \centering
    \begin{subfigure}{.475\linewidth}
        \caption{$X=18$}
        \includegraphics[width=\linewidth,trim=1.5cm 1cm 0cm 0cm]{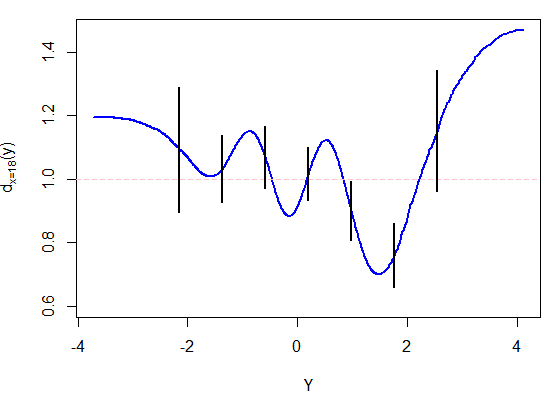}
    \end{subfigure}~
    \begin{subfigure}{.475\linewidth}
        \caption{$X=58$}
        \includegraphics[width=\linewidth,trim=0cm 1cm 1.15cm 0cm]{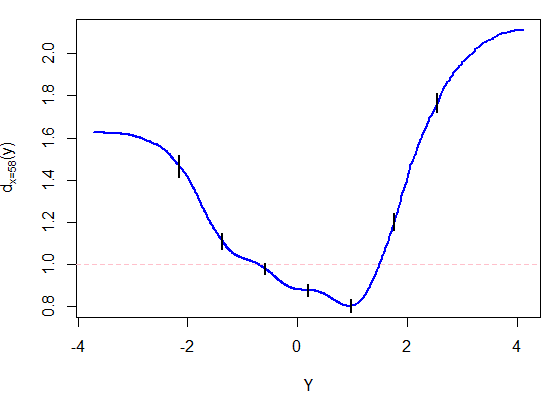}
    \end{subfigure}
    \vskip2em
    \caption{The blue curves are the relevance functions $\whd_x$ at $x=18$ and $x=58$ for the DTI data. The black vertical lines denote $\pm 1\hat{\sigma}_{{\rm boot}}$ error bars.}
    \label{fig:my_label}
    \vskip1em
\end{figure}

The above figure displays the accuracy of the estimated $\whd_x$ at $x=18$ and $x=58$. One noticeable difference between these two figures is the size of the standard error bars, which says $\whd_{x=58}$ is more accurately estimated compared to $\whd_{x=18}$. This is probably because of two reasons: (i) local sample size: we have more samples around $x=58$ than around $x=18$; and (ii) local shape: the shape of $f(y|x=18)$ is considerably more complex relative to the ensemble $f(y)$---which entails higher-order nonlinear customization as reflected in the $\whd_{x=18}$.




\end{document}